\documentclass[lettersize,journal]{IEEEtran}
\IEEEoverridecommandlockouts
\usepackage[T1]{fontenc}
\newif\ifhavebib

\havebibtrue
\usepackage{latexsym}
\usepackage{setspace}
\usepackage{bm}
\usepackage{lipsum}

\newif\ifhavebib

\usepackage{blindtext}
\usepackage{graphicx}
\usepackage{graphicx}
\usepackage{array}
\usepackage{url}
\usepackage{algorithmic}
\usepackage[cmex10]{amsmath}
\usepackage{ifpdf}
\usepackage{amssymb,amsmath}
\usepackage{empheq}
\usepackage{stfloats}   %为了长公式置顶置底
\usepackage{textcomp}
\usepackage{cite}
\usepackage{multirow}
\usepackage{diagbox}
\usepackage{subcaption}
\usepackage{mathrsfs}
\let\oldFootnote\footnote
\newcommand\nextToken\relax

\renewcommand\footnote[1]{%
    \oldFootnote{#1}\futurelet\nextToken\isFootnote}

\newcommand\isFootnote{%
    \ifx\footnote\nextToken\textsuperscript{,}\fi}

\usepackage[ruled,vlined,lined,ruled,boxed]{algorithm2e}
\usepackage[dvipsnames,usenames]{color}
\usepackage[normalem]{ulem}
\usepackage{amsthm}

\usepackage{graphicx}
\usepackage{ifpdf}
\usepackage{cite}
\usepackage{enumerate}
\usepackage{enumitem}

\usepackage{array}
\usepackage{mdwmath}
\usepackage{mdwtab}
\usepackage{eqparbox}
\usepackage{bm}

\usepackage{setspace}
\usepackage{booktabs}
\usepackage[subnum]{cases}
\usepackage{rotating}

\usepackage{listings} 
\definecolor{Red}{rgb}{1,0,0}
\definecolor{Blue}{rgb}{0,0,1}
\definecolor{Olive}{rgb}{0.41,0.55,0.13}
\definecolor{Green}{rgb}{0,1,0}
\definecolor{MGreen}{rgb}{0,0.8,0}
\definecolor{DGreen}{rgb}{0,0.55,0}
\definecolor{Yellow}{rgb}{1,1,0}
\definecolor{Cyan}{rgb}{0,1,1}
\definecolor{Magenta}{rgb}{1,0,1}
\definecolor{Orange}{rgb}{1,.5,0}
\definecolor{Violet}{rgb}{.5,0,.5}
\definecolor{Purple}{rgb}{.75,0,.25}
\definecolor{Brown}{rgb}{.75,.5,.25}
\definecolor{Grey}{rgb}{.5,.5,.5}

\newcommand{\boxhead}[5]{
   \pagestyle{myheadings}
   \thispagestyle{plain}
   \setcounter{page}{1}
   \noindent
   \begin{center}
   \framebox{
      \vbox{\vspace{2mm}
    \hbox to 6.28in { {\bf #1 \hfill} }
       \vspace{6mm}
       \hbox to 6.28in { {\Large \hfill \bf #2  \hfill} }
       \vspace{6mm}
       \hbox to 6.28in { {\it #3 #4 \hfill  #5} }
      \vspace{2mm}}
   }
   \end{center}
   \markboth{#5 -- #2}{#5 -- #2}
   \vspace*{4mm}
}

\usepackage[colorlinks=true, urlcolor=black,  linkcolor=black, citecolor=black]{hyperref}
\theoremstyle{definition}
\newtheorem{theorem}{Theorem} 
\newtheorem{proposition}{Proposition} 
\newtheorem{corollary}{Corollary}

\newtheorem{lemma}{Lemma}

\newtheorem{assumption}{Assumption}
\newtheorem{definition}{Definition}
\newtheorem{example}{Example}
\DeclarePairedDelimiterX{\infdivx}[2]{(}{)}{%
	#1\;\delimsize\|\;#2%
}

\DeclarePairedDelimiter{\norm}{\lVert}{\rVert}
\DeclarePairedDelimiter{\abs}{\lvert}{\rvert}
\DeclareMathOperator*{\argmax}{\mathop{\arg\max}}
\DeclareMathOperator*{\argmin}{\mathop{\arg\min}}
\def\tr{\mathop{\rm tr}\nolimits}%
\def\rank{\mathop{\rm rank}\nolimits}%
\renewcommand{\Pr}{\mathscr{P}}

\newcommand{\ev}{{\bf e}}
\newcommand{\Cv}{{\bf C}}

\newcommand{\Xv}{{\bf X}}

\newcommand{\Uv}{{\bf U}}
\newcommand{\Fv}{{\bf F}}

\newcommand{\rv}{{\bf r}}

\newcommand{\Hv}{{\bf H}}

\newcommand{\Iv}{{\bf I}}
\newcommand{\fv}{{\bf f}}
\newcommand{\gv}{{\bf g}}
\newcommand{\xv}{{\bf x}}

\newcommand{\yv}{{\bf y}}

\newcommand{\zv}{{\bf z}}
\newcommand{\uv}{{\bf u}}
\newcommand{\vv}{{\bf v}}
\newcommand{\hv}{{\bf h}}
\newcommand{\sv}{{\bf s}}

\newcommand{\nv}{{\bf n}}

\newcommand{\zetav}{\boldsymbol \zeta}

\newcommand{\wv}{{\bf w}}

\newcommand{\muv}{\boldsymbol \mu}

\def\e{\epsilon}

\DeclareMathOperator\E{E}

 \def\E{\mathbb{E}}
 
 \def\Pr{\mathrm{Pr}}
\def\de \mathrm{d}
\newcommand{\Norm}{\mathcal{N}}
\newcommand{\CN}{\mathcal{CN}}

\newcommand\eg{e.g.,\xspace}
\newcommand\ie{i.e.,\xspace}
\def\textiid{i.i.d.\@\xspace}

\newcommand\iid{\ifmmode\text{ i.i.d. } \else \textiid \fi}

\newcommand{\Complex}{\mathbb{C}}
\newcommand{\Real}{\mathbb{R}}

\newcommand{\beqs}{\begin{equation*}}
\newcommand{\eeqs}{\end{equation*}}
\newcommand{\beq}{\begin{equation}}
\newcommand{\eeq}{\end{equation}}
\begin{document}

\providecommand{\keywords}[1]{\textbf{\textit{Index terms---}} #1}

\title{Differentially Private Over-the-Air Federated Learning Over MIMO Fading Channels }

\IEEEoverridecommandlockouts

\author{
	Hang Liu,~\IEEEmembership{Member,~IEEE}, Jia~Yan,~\IEEEmembership{Member,~IEEE}, and Ying-Jun~Angela~Zhang,~\IEEEmembership{Fellow,~IEEE}
	\thanks{This work was supported in part by the General Research Fund (project number 14201920, 14202421, 14214122, 14202723), Area of Excellence Scheme grant (project number AoE/E-601/22-R), and NSFC/RGC Collaborative Research Scheme (project number CRS\_HKUST603/22), all from the Research Grants Council of Hong Kong. The work of J. Yan was supported in part by the Guangzhou Municiple Science and Technology Project 2023A03J0011. Part of this work was presented at the  IEEE Global Communications Conference (GLOBECOM), Kuala Lumpur, Malaysia, December 2023 \cite{conference_hljyyz}. 
		
		H. Liu (hl2382@cornell.edu) was with the Department of Information Engineering, The Chinese University of Hong Kong, Hong Kong. He is now with the Department of Electrical and Computer Engineering at Cornell Tech, Cornell University, NY 10044, USA. J. Yan (jasonjiayan@hkust-gz.edu.cn) 
		%was with the Department of Information Engineering, The Chinese University of Hong Kong, Hong Kong. He 
		is with the Intelligent Transportation Thrust, The Hong Kong University of Science and Technology (Guangzhou), China.
		Y.-J. A. Zhang (yjzhang@ie.cuhk.edu.hk) is with the Department of Information Engineering, The Chinese University of Hong Kong, Hong Kong. }
}
%\author{Jia ~Yan$^\dagger$, Suzhi~Bi$^\star$, and Ying-Jun~Angela~Zhang$^\dagger$\\
%$^\dagger$Department of Information Engineering, The Chinese University of Hong Kong, Shatin, N.T., Hong Kong SAR\\
%$^\star$College of Information Engineering, Shenzhen University, Shenzhen, Guangdong, China 518060\\
%E-mail:~ \{yj117, yjzhang\}@ie.cuhk.edu.hk, ~bsz@szu.edu.cn \vspace{-2ex}}

\maketitle
\begin{abstract}
	Federated learning (FL) enables edge devices to collaboratively train machine learning models, with model communication replacing direct data uploading. While over-the-air model aggregation improves communication efficiency, uploading models to an edge server over wireless networks can pose privacy risks. Differential privacy (DP) is a widely used quantitative technique to measure statistical data privacy in FL. Previous research has focused on over-the-air FL with a single-antenna server, leveraging communication noise to enhance user-level DP. This approach achieves the so-called "free DP" by controlling transmit power rather than introducing additional DP-preserving mechanisms at devices, such as adding artificial noise. 
	In this paper, we study differentially private over-the-air FL over a multiple-input multiple-output (MIMO) fading channel. We show that FL model communication with a multiple-antenna server amplifies privacy leakage when the multiple-antenna server employs separate receive combining for model aggregation and information inference. Consequently, relying solely on communication noise, as done in the multiple-input single-output system, cannot meet high privacy requirements, and a device-side privacy-preserving mechanism is necessary for optimal DP design. We analyze the learning convergence and privacy loss of the studied FL system and propose a transceiver design algorithm based on alternating optimization. Numerical results demonstrate that the proposed method achieves a better privacy-learning trade-off compared to prior work.
\end{abstract}


%\IEEEpeerreviewmaketitle
\section{Introduction}
The emergence of artificial intelligence (AI) applications that leverage massive data generated at the edge of wireless networks has attracted widespread interest \cite{8808168,TaskOri6GMag}.  Federate learning (FL) is a popular paradigm for exploiting edge devices' data and computation power for distributed machine learning. FL coordinates the distributive training of an AI model on edge devices by periodically sharing model information with an edge server \cite{FEDSGD}. At the outset, FL appeared to be a promising solution that improves communication efficiency and safeguards data privacy, since on-device data is retained locally and only AI models are communicated. However, recent research indicates that the efficiency of FL is constrained by several challenges, including large communication delays and privacy leakage risks \cite{FL_Challenge}. 

The iterative uploading of high-dimensional local models from massive devices has been identified as the critical bottleneck in FL training. Existing communication standards are shown to be inadequate to support frequent model exchange in FL \cite{GZhu_BroadbandAircomp}. To expedite model uploading, over-the-air computation has been introduced to FL model aggregation. This technique enables devices to concurrently transmit their local gradients by exploiting the superposition property of a multiple-access channel. Pioneering research on over-the-air FL \cite{GZhu_BroadbandAircomp,FL_Digital1} aligned local models at a single-antenna server by channel-inversion-based transmit scaling. Moreover, over-the-air model aggregation was studied with a multiple-antenna receiver to leverage high array gains of the multiple-input multiple-output (MIMO) technique \cite{FL_1}. The design of MIMO over-the-air transceivers requires the joint optimization of transmit scaling and uniform receive beamforming, which can be attained sub-optimally by iterative algorithms, such as semi-definite relaxation \cite{FL_2}, difference-of-convex (DC) \cite{FL_1}, and successive convex approximation \cite{liu2020reconfigurable}.

Apart from the challenges in model communication, the exchange of local models or gradients during FL iterations reveals surprisingly much private information on local data\cite{8737416,8835269,See_grad}. For example, a malicious server can exploit the received gradients to recover sensitive on-device training images \cite{See_grad}. Differential privacy (DP) has been a well-documented quantitative measure of privacy disclosure in machine learning \cite{DP_Book}, which quantifies the probability that an individual data sample can be distinguished from its neighborhoods in a local dataset. 
{Owing to its robust composition theorems, DP has been identified as a powerful tool in dissecting the privacy-learning trade-off in iterative training of FL, thereby simplifying the subsequent system optimization \cite{abadi2016deep,9069945,TMC_DP,9174426}.} 
To maintain DP in FL, local models are typically perturbed by adding random noise before release \cite{abadi2016deep}. In \cite{9069945,TMC_DP}, the authors leveraged Gaussian artificial noise to preserve DP in FL, often referred to as the Gaussian mechanism, where lossless model communication over an orthogonal multiple-access channel was considered. Ref. \cite{9174426} extended the Gaussian mechanism to over-the-air FL and reported a fundamental \emph{learning-privacy trade-off}. Specifically, adding large artificial noise to achieve high privacy levels leads to significant distortion in over-the-air model aggregation, which compromises training convergence. It was further noted in \cite{9174426} that over-the-air model aggregation achieves stronger privacy protection than the orthogonal communication protocol, as artificial noises from all local users are superposed at the receiver.
On the other hand, instead of leveraging additional artificial noise, one can harness inherent communication noise at a single-antenna server to preserve DP, such that arbitrary user privacy levels can be satisfied by solely reducing transmit power \cite{DPFL_DLiu,9322199}.
%Noticeably, Refs. \cite{DPFL_DLiu,9322199} reported that DP can be achieved at a single-antenna server by harnessing communication noise at the receiver. Thus, it is possible to achieve arbitrary user privacy levels at the server by reducing transmit power rather than introducing additional artificial noise. 
Inspired by this, Refs. \cite{DPFL_DLiu,9322199} proposed to transmit power control algorithms that attain accurate FL training while satisfying given DP requirements. Ref. \cite{DPFL_RIS} further balanced the learning-privacy trade-off by reconfigurable intelligent surface. The communication-noise-based DP mechanism was enhanced by incorporating additional perturbations in wireless FL systems, such as inherent anonymity \cite{9413624}, user sub-sampling \cite{9562556}, multi-cell interference \cite{9580422}, hardware-induced distortion \cite{hardware_dp}, and gradient quantization \cite{QUANTIZE_dp}.

The current research on DP-constrained over-the-air FL focuses on model aggregation with a single-antenna server, \ie model uploading over a multiple-input single-output (MISO) channel. The edge devices can preserve DP through optimal power control in local gradient transmission, without the need of injecting artificial noise, even for high DP requirements. This leads to what is called \emph{free DP with zero artificial noise} \cite{DPFL_DLiu}. 
%In this case, edge devices aim to protect their privacy by enforcing DP constraints when sharing local gradients with an untrustworthy server. This setup has an advantage that optimal transmit power control can be achieved without the need of injecting artificial noise, even for high DP requirements, leading to what is called \emph{free DP with zero artificial noise} \cite{DPFL_DLiu}. 
On the other hand, the MIMO technique, which can improve spectral efficiency by deploying a large antenna array at the server, is a crucial feature in current and future wireless protocols \cite{6736761}. MIMO-based over-the-air FL is capable of improving the learning performance by leveraging high array gains to achieve more accurate model aggregation \cite{FL_1}. In this regard, it is necessary to investigate the feasibility of achieving DP in MIMO-based over-the-air FL to accommodate the state-of-the-art network architecture.

However, extending the MISO-based FL design with DP constraints to MIMO systems is not straightforward, due to the coupling effect between the model aggregation protocol and the privacy-preserving mechanism. The corresponding transceiver design challenge is three-fold. First, the privacy leakage in receiving local gradients differs across the server's antennas \cite{Hang_mag}. This makes it possible for an untrustworthy server to exploit separate receive combining strategies for model aggregation and information inference. Designing an optimal information inference mechanism is, however, challenging due to the intractability of privacy loss during the iterative model exchange. Second, model aggregation with the multiple-antenna server is more vulnerable to privacy risks, necessitating a more stringent DP-preserving mechanism on the device side. Third, it is unclear whether the MIMO system can always achieve free DP without inducing artificial noise, as has been verified in the MISO system.

In this work, we study the transceiver design for differentially private over-the-air FL over a MIMO fading channel. Generalizing the model in \cite{9069945,9174426,DPFL_DLiu}, we consider an "honest-but-curious" edge server equipped with a receive antenna array, which aggregates received local models and simultaneously infers local private information of all the devices by multiple receive combiners. We aim to maximize the learning convergence rate while satisfying given DP constraints at the devices.  
%Our work can be taken as a novel extension of the existing research \cite{9069945,9174426,DPFL_DLiu} to MIMO fading channels. 
The contributions of this work are summarized as follows.
\begin{itemize}
	\item We propose a transceiver protocol for MIMO-based differentially private FL. In our design, the devices upload local gradients with a Gaussian DP-preserving mechanism, while the server performs over-the-air model aggregation and parallel private information extraction. We then formulate the transceiver design task as a nested optimization problem. On one hand, the server aims to maximize the privacy loss of each local gradient for private information inference by assuming a fixed transmitter design. On the other hand, we optimize the transmit scaling and over-the-air receive beamforming to maximize the convergence rate while the achieved privacy loss meets the DP constraints.
	\item We analyze the learning convergence rate and the achieved privacy loss of our proposed FL system, under mild assumptions on the training loss function. Based on this analysis, we simplify the system optimization problem with tractable expressions for the convergence rate and the achieved DP of each device.
	\item We demonstrate that the optimization problem exhibits a closed-form optimal solution with a MISO fading channel, which recovers the existing results in \cite{9174426,DPFL_DLiu}. We show that the free-DP property in the MISO setup follows from the inability of the single-antenna receiver to distinguish individual transmitted signals from their superposition. However, we show that this property does not hold in the general MIMO system, and hence the design without artificial noise is sub-optimal in preserving DP for MIMO over-the-air FL.
	\item For the general MIMO system, we derive a closed-form optimal solution to the information extraction optimization at the server. Based on this, we propose an alternating-optimization-based algorithm to sub-optimally solve the transceiver design problem.
\end{itemize}
Our proposed method is evaluated using simulations on both synthetic datasets and real-world image classification tasks, demonstrating its superior performance compared to existing baseline methods. Our results reveal that model aggregation with a MIMO-based server is more vulnerable to privacy breaches than with a single-antenna server. Nonetheless, by jointly optimizing the transceiver protocols, we achieve a better trade-off between learning and privacy due to the more accurate model aggregation through receive beamforming.

The remainder of this paper is organized as follows. In Section \ref{sec2}, we introduce the system model of differentially private over-the-air FL and formulate the system design problem. In Section \ref{sec3}, we analyze the learning convergence rate and the privacy loss to simplify the optimization problem. Then, we study the solutions to the optimization problem for the cases with single-antenna and general multiple-antenna receivers in Sections \ref{sec4} and \ref{sec5}, respectively. In Section \ref{sec6}, we present numerical results to evaluate the proposed method. Finally, this paper concludes in Section \ref{sec7}.

\emph{Notations}: Throughout, 
%we use $\Real$ and $\Complex$ to denote the real and complex number sets, respectively. 
we use regular letters, bold small letters, and bold capital letters to denote scalars, vectors, and matrices, respectively. We use  $\overline \Xv$, $\Xv^T$, and $\Xv^H$ to denote the conjugate, the transpose, and the conjugate transpose of $\Xv$, respectively. We use $\tr(\Xv)$ to denote the trace of $\Xv$ and $\rank(\Xv)$ to denote the rank of $\Xv$.
We use $x_i$ to denote the $i$-th entry of vector $\xv$, $x_{ij}$ to denote the $(i,j)$-th entry of matrix $\Xv$,
% $[\Xv]_{i,:}$ to denote the $i$-th row, and $[\Xv]_{:,j}$ or 
and $\xv_j$ to denote the $j$-th column of $\Xv$. 
The real and circularly-symmetric complex normal distributions with mean $\muv$ and covariance $\Cv$ are denoted by $\Norm(\muv,\Cv)$ and $\CN(\muv,\Cv)$, respectively. The cardinality of set $\mathcal{S}$ is denoted by $\abs{\mathcal{S}}$.
We use  $\norm{\cdot}_p$ to denote the $\ell_p$ norm and $\Iv_N$ to denote the $N\times N$ identity matrix.
%$\bf 1$ to denote the all-one vector with an appropriate size,
%and $\diag(\xv)$ to denote a diagonal matrix with the diagonal entries specified by $\xv$. For any positive integer $n$, we denote the factorial of $n$ by $n!$ and define $[n]\triangleq\{1,2,\cdots,n\}$. 

\section{System Model and Problem Formulation}\label{sec2}
%\subsection{Over-the-Air Federated Learning}

\begin{figure*}[t!]
	\begin{centering}
		\includegraphics[width=6.5 in]{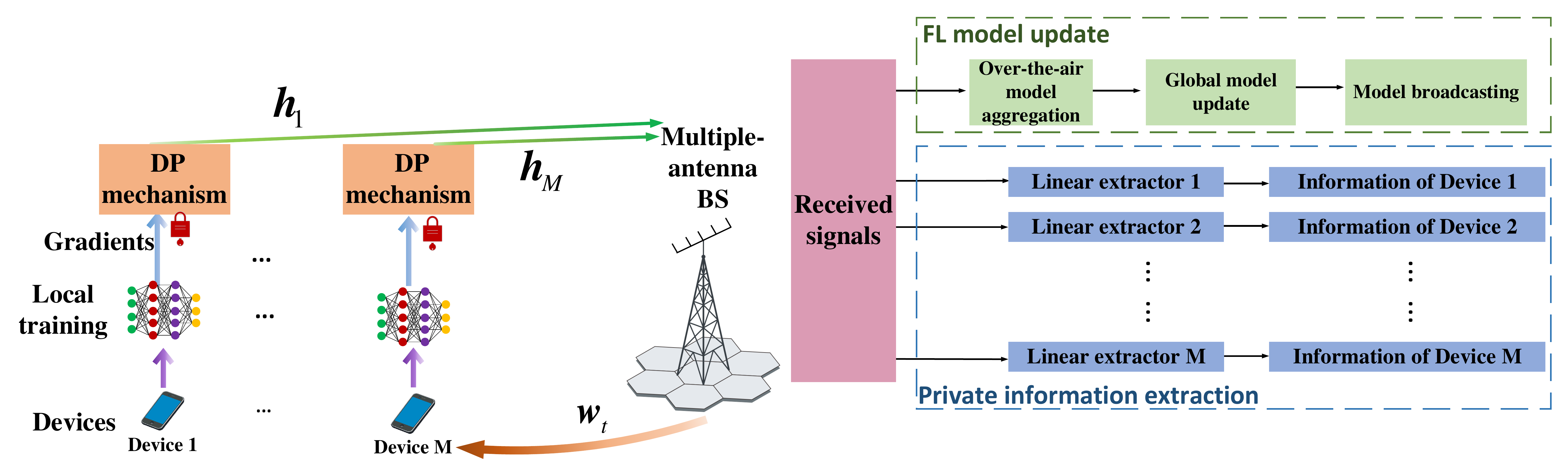}
		\vspace{-0.1cm}
		\caption{The over-the-air FL system with DP mechanisms.}\label{fig1}
	\end{centering}
	\vspace{-0.1cm}
\end{figure*}

We consider an FL system over a wireless network depicted in Fig. \ref{fig1}, where an $N$-antenna base station (BS) acts as a server to coordinate $M$ single-antenna wireless devices (WDs) to train a uniform AI model. The $m$-th WD, $1\leq m\leq M$, has $K_m$ local training samples  with $K\triangleq\sum_{m=1}^MK_m$  denoting the total number of training samples. We denote the local training dataset at the $m$-th WD by $\mathcal{D}_m=\{(\uv_{m,k},v_{m,k});1\leq k\leq K_m\}$, where $\uv_{m,k}$ and $v_{m,k}$ are the input feature vector and the output of the $k$-th training sample, respectively. The FL system aims to minimize an empirical global loss function $F(\cdot)$, as
\begin{align}
	\min_{\wv\in\mathbb{R}^{d\times 1}}F(\wv)&=\frac{1}{K}\sum_{m=1}^{M}\sum_{k=1}^{K_m}f(\wv;(\uv_{m,k},v_{m,k}))\nonumber\\
	&=\frac{1}{K}\sum_{m=1}^{M}K_mF_m(\wv;\mathcal{D}_m),
\end{align}
where $\wv$ signifies the $d$-dimensional model parameter vector, $f(\wv;(\uv_{m,k},v_{m,k}))$ denotes the loss function with respect to (w.r.t.) the sample $(\uv_{m,k},v_{m,k})$, and $F_m(\wv;\mathcal{D}_m)$ is the local loss function at the $m$-th WD defined as
\begin{align}
	F_m(\wv;\mathcal{D}_m)\triangleq\frac{1}{K_m}\sum_{k=1}^{K_m}f(\wv;(\uv_{m,k},v_{m,k})).
\end{align}

The minimization of $F(\wv)$ is performed iteratively with $T$ training rounds by a distributed gradient descent algorithm \cite{BGD}, which requires $T$-round model parameter communications between the BS and the WDs through wireless fading channels. Denote the channel coefficient vector between the BS and the $m$-th WD by $\hv_m\in\Complex^{N\times 1}$. We assume that the channel coefficients $\{\hv_m\}_{m=1}^M$ are invariant in $T$ rounds and the perfect channel state information (CSI) is available by following \cite{FL_1, GZhu_BroadbandAircomp,liu2020reconfigurable}.\footnote{The proposed FL design in this work can be readily extended to time-varying channels by following \cite{liu2020reconfigurable}. Specifically, the system designing variables can be updated by revoking the proposed optimization algorithm when the instantaneous channel coefficients change.}

The FL training process is discussed as follows.
In the $t$-th iteration, $1\leq t\leq T$, the BS first broadcasts the current global model $\wv_t$ to the $M$ WDs through noiseless downlink channels. Accordingly, each WD computes the local gradient of $F_m(\wv;\mathcal{D}_m)$ at $\wv=\wv_t$ as
\begin{align}\label{eq03}
	\mathbf{g}_{m,t}=\frac{1}{K_m}\sum_{k=1}^{K_m}\nabla f(\wv_t;(\uv_{m,k},v_{m,k})), 1\leq  m \leq M,
\end{align}
where $\nabla f(\wv_t;(\uv_{m,k},v_{m,k}))$ is the gradient of $f(\wv;(\uv_{m,k},v_{m,k}))$ at $\wv=\wv_t$.
We make the following assumption on the gradient norm, which has been widely adopted in the FL literature.
\assumption{\label{assump1}The (scaled) $\ell_2$-norm of each local gradient $\nabla f(\cdot)$ is upper bounded by a known constant $L>0$, \ie
	\begin{align}\label{eq07}
		\frac{1}{\sqrt d}	\|\nabla f(\wv_t;\uv_{m,k},v_{m,k})\|_2\leq L, \forall m,k,t.
	\end{align}
}
Assumption \ref{assump1} can be satisfied by using gradient clipping in practice. Specifically, with a given $L$, each $\nabla f(\cdot)$ is replaced with $\nabla f(\cdot)/{\max\{1,\norm{\nabla f(\cdot)}_2/(\sqrt d L)\}}$ to ensure \eqref{eq07}. We note that such a clipping method is popular in FL training for the purpose of enhancing learning convergence \cite{Communicate_Little} and facilitating DP design \cite{TMC_DP, DPFL_DLiu}.  
Assumption \ref{assump1}  implies a bound of the gradient vector $\gv_{m,t}$ in \eqref{eq03} by applying the triangle inequality, \ie  
\begin{align}\label{eq5}
	\frac{1}{\sqrt d}	\norm{\gv_{m,t}}_2
	%&=\frac{1}{\sqrt dK_m}\big\|\sum_{k=1}^{K_m}\nabla f(\wv_t;(\uv_{m,k},v_{m,k}))\big\|_2
	\leq \frac{1}{\sqrt dK_m}\sum_{k=1}^{K_m}\|\nabla f(\wv_t;(\uv_{m,k},v_{m,k}))\|_2\leq L.
\end{align}

After computing the gradient vector, each WD transforms $\gv_{m,t}$ to a  $d$-dimensional signal vector $\xv_{m,t}$ by using a DP-enhancing mapping function $\mathcal{M}_m(\cdot):\Real^{d\times 1}\to \Complex^{d\times 1}$. The detail of $\mathcal{M}_m(\cdot)$ will be discussed in Section \ref{sec_2a}. The signals at all the WDs $\{\xv_{m,t}:1\leq m\leq M\}$ are transmitted to the BS through uplink channels. Then, in order to estimate the global gradient for the global model update, the BS performs model aggregation by computing a weighted sum of the local gradient vectors from the received signals. Denote the true global gradient vector by $\mathbf{g}_t\triangleq \sum_{m=1}^MK_m\mathbf{g}_{m,t}$ and its estimate at the BS by $\hat \gv_t$. After obtaining $\hat \gv_t$, the BS updates the global model $\wv_{t+1}$ with a predefined constant learning rate $\lambda>0$ as
\begin{align}\label{global_update}
	\wv_{t+1}=\wv_{t}-\frac{\lambda}{K}\hat{\mathbf{g}}_t.
\end{align}
%where $\lambda$ is the learning rate. 
%The above steps are repeated until we finish the $T$-th training iteration.

It is shown that the uplink model aggregation is the bottleneck limiting the performance of FL \cite{GZhu_BroadbandAircomp}. 
%The impact of the aggregation step in \eqref{global_update} is two-fold: First,
Specifically, the communication noise and wireless fading cause inevitable estimation error in $\hat \gv_t$ and consequently slow down the training convergence. On the other hand, the transmissions of the local gradients leak sensitive information about local datasets, which can be used by an untrusted server to recover local training samples \cite{See_grad} or infer their private features \cite{8737416,8835269}. The DP-preserving mechanism  $\mathcal{M}_m$ can be designed to preserve local privacy by perturbing local gradients, which, however, degrades the model aggregation accuracy and exacerbates the convergence. Therefore, it is of great importance to optimize the model transmission protocol for achieving efficient model aggregation while protecting local private information.

In this work, we consider the threat model with an "honest-but-curious" BS and trustworthy WDs \cite{TMC_DP,DPFL_DLiu}; see Fig. \ref{fig1}. We sketch the workflow of the network nodes here and discuss the detailed protocols in the subsequent sections.
On one hand, the BS follows the aforementioned model aggregation policy to estimate $\hat\gv_t$ and update the global model by \eqref{global_update}. Meanwhile, the BS is curious about the information of \emph{all} the local datasets. It adopts parallel information extractors to estimate $M$ vulnerable vectors $\{\rv_{m,t}\}_{\forall m}$ from the received signals, one for each local dataset, 
% $\{\rv_{m,t}\in\Complex^{d\times 1}:1\leq m\leq M\}$ 
and attempts to infer private features from them.
% he received signal model and the information extraction process will be detailed in Section \ref{sec_2b}.
On the other hand, each WD computes and uploads local gradients to the BS without curiosity about other WDs' private information. In order to protect the private information of its own dataset, each WD uses a privacy-preserving mechanism to mask its local gradients before uploading. 

Here, we assume perfect downlink broadcasting owing to the adequate transmission power budget at the BS and focus on the uplink model aggregation design as in \cite{GZhu_BroadbandAircomp, FL_1,liu2020reconfigurable, DPFL_DLiu, DPFL_RIS}.\footnote{Downlink model broadcasting may also be prone to information leakage. For example, an external eavesdropper can intercept model information from the downlink channels. Since this work focuses on \emph{uplink model aggregation} design, secure downlink model broadcasting is beyond the scope of this paper and is left for future explorations.} In the sequel, we leverage \emph{analog} over-the-air computation to achieve fast uplink model aggregation, along with a DP mechanism to guarantee local data privacy. The following two sections discuss the communication protocols at the WDs and the BS, respectively.  

\subsection{Transmitting Protocol at the WDs: Over-the-Air Model Uploading With DP}\label{sec_2a}

After computing the local gradient $\gv_{m,t}$ in \eqref{eq03}, the $m$-th WD, $1\leq m\leq M$, adopts a DP mechanism  $\mathcal{M}_m(\cdot)$ to encode $\gv_{m,t}$ to $\xv_{m,t}\in\Complex^{d\times 1}$, i.e., the transmitted signal to the BS.

\subsubsection{Preliminaries on DP}
To explain the DP mechanism, we begin by presenting the definition and measurement of user privacy used in this work. Consider any two possible \emph{adjacent datasets} in the $m$-th WD, denoted by $\mathcal{D}_m$ and $\mathcal{D}_m^\prime$, that differ only one training sample with each other with the same cardinality $|\mathcal{D}_m|=|\mathcal{D}_m^\prime|=K_m$. Suppose that the $m$-th WD uses either the dataset $\mathcal{D}_m$ or $\mathcal{D}_m^\prime$ for local training with the datasets of the other $M-1$ WDs fixed. In any $t$-th round, the $m$-th WD computes $\gv_{m,t}$ by  \eqref{eq03} w.r.t. either $\mathcal{D}_m$ or $\mathcal{D}_m^\prime$, applies $\mathcal{M}_m(\cdot)$ to transform $\gv_{m,t}$ to $\xv_{m,t}$, and uploads $\xv_{m,t}$ to the BS. Then, the BS processes the received signals with a linear information extractor for each WD $m$ and outputs a vector $\rv_{m,t}$ that is vulnerable to leak information on the local dataset. 
We denote the collection of $\rv_{m,t}$ obtained in $T$ rounds  by ${\bf \mathcal{R}}_m\triangleq\{\rv_{m,t}:1\leq t\leq T\}$. 
%Similarly, if the $m$-th device uses $\mathcal{D}_m^\prime$ for local model training with the other WDs' datasets unchanged, we denote the resultant collection of the received signals in the whole $T$ iterations by $\yv^\prime$. 
The following statement defines \emph{local differential privacy} \cite{DP_Book}.
\definition{For any given $\epsilon_m>0$, $\delta_m\in [0,1]$, and any two possible adjacent datasets $\mathcal{D}_m$ and $\mathcal{D}_m^\prime$ of the $m$-th WD, the FL system after $T$ training rounds 
	%	 mapping function $\mathcal{M}_m(\cdot)\in\Real^{d\times 1}\to \Complex^{d\times 1}$ that generates the transmitting signals of the $m$-th WD is said to 
	satisfies the  $(\epsilon_m,\delta_m)$-DP for the $m$-th WD with mechanism $\mathcal{M}_m(\cdot)$  if the following inequality holds
	\begin{align}\label{eq05}
		\Pr(\mathcal{R}_m|\mathcal{D}_m)\leq e^{\epsilon_m}\Pr(\mathcal{R}_m|\mathcal{D}_m^\prime)+\delta_m.
	\end{align}\label{def1}
}
As shown in \cite{DP_Book}, the $(\epsilon_m,\delta_m)$-DP in \eqref{eq05} implies a concentration bound on the log-likelihood ratio of distinguishing the adjacent datasets $\mathcal{D}_m$ and $\mathcal{D}_m^\prime$ by observing ${\bf \mathcal{R}}_m$, \ie
\begin{small}
	\begin{align}\label{eq06}
		&\Pr\left(\left|\ln\frac{\Pr({\bf \mathcal{R}}_m|\mathcal{D}_m)}{\Pr({\bf \mathcal{R}}_m |\mathcal{D}_m^\prime)}\right|\leq \epsilon_m\right)\nonumber\\
		&=\Pr\left(\left|\sum_{t=1}^T\ln\left(\frac{\Pr(\rv_{m,t}|\rv_{m,t-1},...,\rv_{m,1};\mathcal{D}_m)}{\Pr(\rv_{m,t}|\rv_{m,t-1},...,\rv_{m,1};\mathcal{D}_m^\prime)}\right)\right|\leq \epsilon_m\right)\geq 1-\delta_m. 
	\end{align}
\end{small}
{As indicated by \eqref{eq06}, when the server employs the log-likelihood ratio test over the observation ${\bf \mathcal{R}}_m$, there is a high probability that it fails to distinguish between $\mathcal{D}_m$ and $\mathcal{D}_m^\prime$ with small $\epsilon_m$ and $\delta_m$.}
Denote the privacy loss at the $t$-th training round as $\mathcal{L}_{\mathcal{D}_m,\mathcal{D}_m^\prime}(\rv_{m,t})\triangleq\ln\frac{\Pr(\rv_{m,t}|\rv_{m,t-1},...,\rv_{m,1};\mathcal{D}_m)}{\Pr(\rv_{m,t}|\rv_{m,t-1},...,\rv_{m,1};\mathcal{D}_m^\prime)}$. From \eqref{eq06}, we see that $\epsilon_m$ probabilistically bounds the \emph{accumulated privacy loss} as the FL training progresses, i.e., $\Pr\left(\left|\sum_{t=1}^T\mathcal{L}_{\mathcal{D}_m,\mathcal{D}_m^\prime}(\rv_{m,t})\right|\leq \epsilon_m\right)\geq 1-\delta_m$.
%the probability that the  true dataset is distinguishable at the BS from any possible adjacent candidate. 
As such, identifying the data samples in $\mathcal{D}_m$ is almost impossible (i.e., small accumulated privacy loss) with sufficiently small $\delta_m$ and $\e_m$.
%We see from \eqref{eq06} that 
Accordingly, \emph{a smaller $\epsilon_m$ means a more private FL training process} for the $m$-th WD with a fixed $\delta_m$.

\subsubsection{DP-based Transmit Signal Design}
We consider that the $m$-th WD, $1\leq m\leq M$, imposes a constraint on the $(\epsilon_m^{(\text{WD})},\delta_m)$-DP for protecting its privacy, where $\delta_m$ is a predefined hyper-parameter shared with the BS, and $\epsilon_m^{(\text{WD})}$ represents the \emph{desired DP requirement} at the $m$-th WD.
To fulfill the $(\epsilon_m^{(\text{WD})},\delta_m)$-DP requirement, each WD $m$ adopts the \emph{Gaussian mechanism} to design $\mathcal{M}_m$  by adding Gaussian-distributed artificial noise to the input $\gv_{m,t}$. Specifically, the transmitting signal vector of the $m$-th WD in the $t$-th round is  set to
\begin{align}\label{eq08}
	\xv_{m,t}= \mathcal{M}_m(\gv_{m,t})=\frac{s_{m,1}}{L}\gv_{m,t}+s_{m,2}\nv_{m,t}, \forall m,t,
\end{align}
where $\nv_{m,t}\in\Complex^{d\times 1}$ is a $d$-dimensional artificial noise vector following the distribution of $\Norm({\bf 0},\Iv_d)$, and $s_{m,1}\in\Complex$  (or $s_{m,2}\in\Complex$) is the transmit scalar for the gradient vector (or the artificial noise). 

The decision variables $s_{m,1}$ and $s_{m,2}$ control the power fractions of the gradient vector and the artificial noise, respectively, which reflects the trade-off between FL training and DP. We shall optimize $\{s_{m,1},s_{m,2}\}$ to fulfill the model aggregation and local DP requirements under the following   transmit power constraint:
\begin{align}\label{eq09}
	\frac{1}{d}\E[\norm{\xv_{m,t}}_2^2]=|s_{m,1}|^2\frac{\norm{\gv_{m,t}}_2^2}{d L^2}+|s_{m,2}|^2\leq P_{\text{max}},\forall m,t,
\end{align}
where $P_{\text{max}}$ is the maximum transmit power per symbol for each WD. Since we have $\norm{\gv_{m,t}}_2^2\leq dL^2$ from \eqref{eq5}, the condition in \eqref{eq09} can be guaranteed if
\begin{align}\label{eq10}
	|s_{m,1}|^2+|s_{m,2}|^2\leq P_{\text{max}},\forall m,t.
\end{align}

We adopt over-the-air computation to efficiently upload $\{\xv_{m,t}\}_{m=1}^M$ by exploiting 
the super-position property of wireless multiple-access channels. In each training iteration, the $M$ signal vectors $\{\xv_{m,t}\}$ are transmitted simultaneously entry by entry through the same radio resources with $d$ time slots needed, where the $i$-th entries $\{x_{m,t}[i]\}_{m=1}^M$ are transmitted in the $i$-th slot for $1\leq i\leq d$. Consequently, a large number of WDs can concurrently upload their signals with low latency as the required radio resources are independent of the number of WDs.  The received signal model is detailed in what follows.

\subsection{Receiving Protocol at the BS: Model Aggregation and Private Information Extraction}\label{sec_2b}

\subsubsection{Model Aggregation}
With over-the-air model aggregation, the transmitted signals are naturally superposed at the BS. Accordingly, the received signal vector at the BS in the $i$-th time slot of the $t$-th training round, denoted by $\yv_t[i]\in\Complex^{N\times 1}$, is given by
\begin{align}\label{eq11}
	&	\mathbf{y}_{t}[i]=\sum_{m=1}^M\hv_{m}x_{m,t}[i]+\zv_t[i]\nonumber\\
	&=\sum_{m=1}^M\hv_{m}\left(\frac{s_{m,1}}{L}g_{m,t}[i]+s_{m,2}n_{m,t}[i]\right)+\zv_t[i],\forall i,
\end{align}
where $g_{m,t}[i]$ and $n_{m,t}[i]$ are the $i$-th entries of $\gv_{m,t}$ and $\nv_{m,t}$, respectively, and $\zv_t[i]\in\mathbb{C}^{N\times 1}$ is an additive white Gaussian noise (AWGN) following the distribution of $\CN({\bf 0},\sigma_z^2\Iv_N)$.

In the $i$-th time slot, the BS computes the estimate of $g_t[i]=\sum_{m=1}^MK_mg_{m,t}[i]$, i.e., $\hat{g}_t[i]$, by using a linear combining technique, \ie
\begin{align}\label{gradFL}
	&\hat{g}_t[i]=\frac{1}{\sqrt{\eta}}\fv_0^H\mathbf{y}_{t}[i]\nonumber\\
	&=\sum_{m=1}^M\frac{1}{\sqrt{\eta}}\fv_0^H\hv_{m}\frac{s_{m,1}}{L}g_{m,t}[i]\nonumber\\
	&~~~~+\frac{1}{\sqrt{\eta}}\left(\sum_{m=1}^M\fv_0^H\hv_ms_{m,2}n_{m,t}[i]+\fv_0^H\mathbf{z}_t[i]\right),
\end{align}
where $\fv_0\in\Complex^{N\times 1}$ is the unit-norm receive combiner with $\|\fv_0\|^2_2=1$, and $\eta>0$ is the normalizing factor. After estimating $\hat{\mathbf{g}}_t=\{\hat{g}_t[1],...,\hat{g}_t[d]\}$, the BS updates the global model by \eqref{global_update}.

\subsubsection{Private Information Extraction}
Besides estimating $\hat\gv_t$  for global model update, the "honest-but-curious" BS also intends to infer private information on local datasets from the received signals. {While model aggregation focuses on enhancing the aggregation accuracy of the global gradients, privacy information extractors maximize the privacy loss of each local dataset.} Due to the DP mechanisms at the transmitters and limited degrees of freedom in the processor $\{\fv_0,\eta\}$ in \eqref{gradFL}, it is challenging to rely solely on $\hat\gv_t$ to extract private information of all the $M$ local datasets.
%Due to the DP mechanism at the transmitters, it is difficult to decipher information directly from $\hat \gv_t$.
{The decomposable nature of the privacy loss, when evaluated on each specific dataset, prompts our design to utilize $M$ parallel linear estimators $\fv_1,\cdots, \fv_M$ to process $\yv_t[i]$}. The output of the $m$-th estimator is given by
\begin{align}\label{eq13}
	&r_{m,t}[i]=\fv_m^H\mathbf{y}_{t}[i]\nonumber\\
	&=\sum_{m^\prime=1}^M\fv_m^H\hv_{m^\prime}\frac{s_{m^\prime,1}}{L}g_{m^\prime,t}[i]\nonumber\\
	&~~~~+\sum_{m^\prime=1}^M\fv_m^H\hv_{m^\prime}s_{m^\prime,2}n_{m^\prime,t}[i]+\fv_m^H\zv_t[i],
\end{align}
where $\fv_m$ is the $m$-th information-extraction estimator and is assumed to have a unit norm $\norm{\fv_m}_2=1$ without loss of generality. Denote $\rv_{m,t}=[r_{m,t}[1],\cdots,r_{m,t}[d]]\in\Complex^{d\times 1}$.  Each information extractor $m$ at the BS targets at designing $\fv_m$, so that ${\bf \mathcal{R}}_m=\{\rv_{m,t}:1\leq t\leq T\}$ leaks as much information on the $m$-th dataset $\mathcal{D}_m$ as possible (i.e., large accumulated privacy loss).
%than $\{\hat \gv_t\}_{t=1}^T$ after $T$ training rounds.

%In other words, 
%As a consequence, with the BS-side information extraction mechanism in \eqref{eq14}, \emph{the $m$-th WD can achieve the $(\epsilon_m^{(\text{BS})}(\fv_m^\star,\{s_{m,1}, s_{m,2}\}_{m=1}^M),\delta_m)$-DP at best}. In other words, the FL system only achieves 

\subsection{Problem Formulation}
\subsubsection{Optimal Private Information Extraction}
The goal of each private information extractor at the BS is to maximize the accumulated privacy loss of WD $m$'s local dataset w.r.t. the disclosed signals ${\bf \mathcal{R}}_m$ in \eqref{eq13}.
Leveraging the DP property in \eqref{eq06},
we denote the probabilistic bound of the accumulated privacy  at the BS by $\epsilon_m^{(\text{BS})}(\fv_m,\{s_{m,1}, s_{m,2}\}_{m=1}^M)$, i.e.,
%is lower than that of $\hat \gv_t$ in \eqref{gradFL}.   
%With a slight abuse of notation, 
%, the overall privacy loss regarding $\mathcal{D}_m$ associated with a given $\delta_m\in [0,1]$ is bounded by , i.e.,
\begin{align}\label{achievedDP}
	\Pr\left(\left|\sum_{t=1}^T\mathcal{L}_{\mathcal{D}_m,\mathcal{D}_m^\prime}(\rv_{m,t})\right|\leq \e_m^{(\text{BS})}(\cdot)\right)\geq 1-\delta_m, \forall \mathcal{D}_m,\mathcal{D}_m^\prime, m,
\end{align}
where $\delta_m$ is given by the WD $m$. We note that $\epsilon_m^{(\text{BS})}$ is the function value measuring the \emph{achieved} DP at the BS, depending on the transmit variables $\{s_{m,1}, s_{m,2}\}$ and the receive vector $\fv_m$. Specifically, a larger  $\e_m^{(\text{BS})}(\cdot)$ means more privacy leakage of the uplink model aggregation, \ie a higher privacy loss.
Accordingly, for any given $\{s_{m,1}, s_{m,2}\}_{m=1}^M$ at the transmitter, the BS optimizes $\fv_m$ to maximize $\epsilon_m^{(\text{BS})}(\cdot)$ as
\begin{align}\label{eq14}
	\fv^\star_m
	%	\left(\{s_{m,1},s_{m,2}\}_{m=1}^M\right)
	=\argmax_{\norm{\fv_m}_2^2=1}
	\epsilon_m^{(\text{BS})}(\fv_m,\{s_{m,1}, s_{m,2}\}_{m=1}^M), \forall m.
\end{align}
%where we explicitly indicate that the desired vector $\fv_m^\star(\cdot)$ is a function of $\{s_{m,1}, s_{m,2}\}_{m=1}^M$. 
With the optimal $\fv^\star_m$ above, the $(\epsilon_m^{(\text{BS})}(\cdot),\delta_m)$-DP can be achieved at the BS w.r.t. WD $m$'s local dataset.
%Accordingly, the \emph{achievable DP level at the BS} w.r.t. WD $m$'s local dataset is given by the objective value in \eqref{eq14} as $\epsilon_m^{(\text{BS})}(\fv_m^\star,\{s_{m,1}, s_{m,2}\}_{m=1}^M)$. %We note that this is the best possible DP that can be achieved by the $m$-th WD. 

\subsubsection{Optimal FL Performance with DP}
%Given that each WD $m$ intends to guarantee the $(\epsilon_m^{(\text{WD})},\delta_m)$-DP, we need to ensure its achievability at the BS. In other words, we
With the optimal private information extractors $\{\fv^\star_m\}$ in \eqref{eq14}, in order to guarantee the $(\epsilon_m^{(\text{WD})},\delta_m)$-DP for WD $m$, one should optimize $\{s_{m,1}, s_{m,2}\}_{m=1}^M$ such that\footnote{Notice that $\epsilon_m^{(\text{WD})}$ is the pre-defined constant representing the \emph{desired} DP requirement of WD $m$, while $\epsilon_m^{(\text{BS})}$ is the \emph{achieved} DP level at the BS.}
\begin{align}
	\epsilon_m^{(\text{BS})}(\fv_m^\star,\{s_{m,1}, s_{m,2}\}_{m=1}^M) \leq \epsilon_m^{(\text{WD})},\forall m.
\end{align}
%$\epsilon_m^{(\text{BS})}(\fv_m^\star,\{s_{m,1}, s_{m,2}\}_{m=1}^M)$ $\leq \epsilon_m^{(\text{WD})},\forall m$.  
Moreover, since the over-the-air model aggregation accuracy critically affects the FL training convergence \cite{liu2020reconfigurable}, we minimize the expected training loss after $T$ training rounds by optimizing  the transceiver designing variables  $\{\eta,\fv_0,\{s_{m,1},s_{m,2}\}_{m=1}^{M}\}$, yielding the following optimization problem:
\begin{subequations}
	\begin{align}
		(\text{P}1):	&\min_{\eta,\fv_0,\{s_{m,1},s_{m,2}\}_{m=1}^{M}}~~ \E[F(\wv_T)]\label{P1a}\\
		&\text{s.t. } 
		%& \Pr(\rv_m|\mathcal{D}_m)\leq e^{\epsilon_m^{(\text{WD})}}\Pr(\rv_m|\mathcal{D}_m^\prime)+\delta_m, \forall \mathcal{D}_m,\mathcal{D}_m^\prime, \forall m, \label{P1b}\\
		\epsilon_m^{(\text{BS})}(\fv_m^\star,\{s_{m,1}, s_{m,2}\}_{m=1}^M)\leq \epsilon_m^{(\text{WD})},\forall m,\label{P1c}\\
		& ~~~~|s_{m,1}|^2+|s_{m,2}|^2\leq P_{\text{max}},\forall m,\label{P1d}\\
		&~~~~ \norm{\fv_0}_2^2=1, \eta>0,\label{P1e}
	\end{align}
\end{subequations}
where the expectation in \eqref{P1a} is taken w.r.t. the communication noise, %$\fv_m^\star$ in \eqref{P1c} is given in \eqref{eq14}, 
\eqref{P1c} represents the DP requirement at each WD, 
%\eqref{P1c} ensures the feasibility of the DP constraint, 
and \eqref{P1d}--\eqref{P1e} are the power constraints of the designing variables. 

Problem (P1) is challenging to solve. %due to the following reasons. 
First, the impact of transceiver design on model aggregation error is accumulated in training iterations, making the expression of the expected loss $\E[F(\wv_T)]$ highly intractable. 
Second, it is difficult to obtain the analytical expression of $\epsilon_m^{(\text{BS})}(\cdot)$ in \eqref{achievedDP} as it accounts $T$-round private information leakage.
%Second, directly evaluating the DP constraint in \eqref{achievedDP} is computationally prohibitive even for fixed variables. 
Finally, (P1) is a nested optimization problem, where evaluating $\epsilon_m^{(\text{BS})}$ in  \eqref{P1c} for fixed $\{s_{m,1}, s_{m,2}\}_{m=1}^M$ involves solving $M$ non-convex sub-problems in \eqref{eq14}. 
To tackle these challenges, we shall pursue an approximate problem to (P1) by minimizing an upper bound of \eqref{P1a} and deriving a closed-form expression of $\epsilon_m^{(\text{BS})}(\fv_m,\{s_{m,1}, s_{m,2}\}_{m=1}^M)$ for \eqref{P1c}.
%, as detailed in the next section.   
%tightening the constraints \eqref{P1b}--\eqref{P1c}, as detailed in the next section.

\section{Performance Analysis and Problem Simplification}\label{sec3}
In this section, a tractable approximation is constructed for Problem (P1). We start by deriving a closed-form upper bound for the expected training loss $\E[F(\wv_T)]$ under several assumptions on the training model $F(\cdot)$.
\subsection{Learning Convergence Analysis}
First, we list three assumptions imposed on the training loss function.
\assumption{\label{assump2} $F(\cdot)$ is $\mu$-strongly convex for some $\mu>0$, \ie for $ \forall\wv,\wv'\in\mathbb{R}^{d\times 1}$,
	\begin{align}
		F(\wv)\geq F(\wv')+(\wv-\wv')^T\nabla F(\wv')+\frac{\mu}{2}\|\wv-\wv'\|_2^2.
	\end{align}
}
Assumption \ref{assump2} guarantees that a global minimum $\wv^\star$ exists for the training loss function $F(\cdot)$.

\assumption{\label{assump3}The gradient $\nabla F(\cdot)$ is Lipschitz continuous for some $\omega>0$, \ie
	\begin{align}
		\|\nabla F(\wv)-\nabla F(\wv')\|_2\leq\omega\|\wv-\wv'\|_2, \forall\wv,\wv'\in\mathbb{R}^{d\times 1}.
	\end{align}
}
\assumption{\label{assump4} $F(\cdot)$ is twice-continuously differentiable.
}

{Assumptions \ref{assump2} and \ref{assump4} are satisfied when the Hessian matrix of $F(\cdot)$ is continuous and positive definite.}
We emphasize that the above assumptions, together with Assumption \ref{assump1} in Section \ref{sec2}, are standard in the stochastic optimization literature, such as \cite{Friedlander2012, NDP, CFLIT}, to name a few. Typical examples satisfying  Assumptions 2--4 are regularized problems, such as ridge regression, $\ell_2$-norm-regularized logistic regression, and  support vector machine \cite{DL_Goodfellow}. 
Under Assumptions 1--4, we derive a closed-form upper bound for the training loss as follows.
\theorem{\label{theorem1}Suppose $F(\cdot)$ satisfies Assumptions 1-4. With the learning rate $\lambda=\frac{1}{\omega}$, we have
	\begin{align}\label{upperbound}
		&\mathbb{E}[F(\wv_{T})]\leq F(\wv^\star)+ B^{T}(F(\wv_{0})-F(\wv^\star))\nonumber\\ 
		&+A(\{s_{m,2}\},\eta,\fv_0)\frac{1-B^{T}}{1-B}
		+\sum_{t=1}^TB^{T-t}{C_{t}}(\{s_{m,1}\},\eta,\fv_0),
	\end{align}
	where $\wv^\star$ is the minimum of $F(\cdot)$, and the expectation is taken w.r.t. the communication noise. Here, $B=1-{\mu}/{\omega}$, and  $A(\cdot)$ and $C_t(\cdot)$ are auxiliary functions defined as 
	\begin{align}\label{A}
		A(\{s_{m,2}\},\eta,\fv_0)&\triangleq\frac{d\left(\sum_{m=1}^M|\fv_0^H\hv_m|^2|s_{m,2}|^2+\sigma_z^2\right)}{2\omega K^2\eta},\\
		C_{t}(\{s_{m,1}\},\eta,\fv_0)&\triangleq\frac{\sum_{i=1}^d\mathbb{E}\bigg[\left|\sum_{m}\left(K_m-\frac{\fv_0^H\hv_ms_{m,1}}{\sqrt{\eta}L}\right)g_{m,t}[i]\right|^2\bigg]}{2\omega K^2}.\label{C}
	\end{align}
}
\begin{IEEEproof}
	See Appendix \ref{appa}.
\end{IEEEproof}

The second term on the right-hand side (r.h.s.) of \eqref{upperbound} denotes the loss residue from initialization, which decreases exponentially with the training round $T$. Moreover, the artificial noise induced by the DP mechanism, along with the communication noise, leads to a non-diminishing loss in the third term of \eqref{upperbound}. Finally, the last term of \eqref{upperbound} corresponds to the potential signal mismatch in over-the-air model aggregation.

\subsection{DP Analysis}
%Although the DP expression in \eqref{achievedDP} is intractable,
Due to the random noise perturbation in the DP mechanism as well as the composition of privacy losses, it is challenging to bound the overall privacy losses in \eqref{achievedDP} w.r.t. $\{\rv_{m,t}:1\leq t\leq T\}$. To tackle this challenge, we first decouple $r_{m,t}[i]$ in \eqref{eq13} as
\begin{align}
	r_{m,t}[i]=\fv_m^H\hv_{m}\frac{s_{m,1}}{L}g_{m,t}[i]+\sum_{m^\prime\neq m}\fv_m^H\hv_{m^\prime}\frac{s_{m^\prime,1}}{L}g_{m^\prime,t}[i]+q_{t}[i],
\end{align}
where the first term is the disclosed signal regarding dataset $\mathcal{D}_m$ of WD $m$; the second term is the inter-user interference; and the third term $q_{t}[i]\triangleq\sum_{m^\prime=1}^M\fv_m^H\hv_{m^\prime}s_{m^\prime,2}n_{m^\prime,t}[i]+\fv_m^H\zv_t[i]$ is the Gaussian noise introduced by the DP mechanisms and the communication noise. Here, the Gaussian noise is superposed to protect all the local gradients with over-the-air computation.
%all revealed signals from WDs are protected by the same superposed Guassian noise $q_{t}[i]$.

We derive a closed-form upper-bound $\e_m^{(\text{BS})}(\cdot)$ of the accumulated privacy loss by following \cite{abadi2016deep,TMC_DP}. Specifically, $\e_m^{(\text{BS})}(\cdot)$ depends on the sensitivity of the noise-free disclosed signal regarding the input dataset of WD $m$. Associated with two adjacent datasets $\mathcal{D}_m$ and $\mathcal{D}_m^\prime$ of WD $m$, the revealed signals at training round $t$ are $\fv_m^H\hv_m\frac{s_{m,1}}{L}\mathbf{g}_{m,t}(\mathcal{D}_m;\cdot)$ and $\fv_m^H\hv_m\frac{s_{m,1}}{L}\mathbf{g}_{m,t}(\mathcal{D}'_m;\cdot)$, respectively, where $\mathbf{g}_{m,t}(\mathcal{D}_m;\cdot)$ denotes the local gradient vector computed w.r.t. $\mathcal{D}_m$. Accordingly, the $l_2$-sensitivity at training round $t$ bounds the output perturbation of the DP mechanism, given by the maximum difference of the revealed signals, i.e.,
\begin{align}
	\Delta_{m,t}\triangleq \max_{\mathcal{D}_m,\mathcal{D}'_m}\big\|\fv_m^H\hv_m\frac{s_{m,1}}{L}\left(\mathbf{g}_{m,t}(\mathcal{D}_m;\cdot)-\mathbf{g}_{m,t}(\mathcal{D}'_m;\cdot)\right)\big\|_2.
\end{align}
The maximum sensitivity of the $m$-th WD's dataset in $T$ training rounds is given by $\Delta_m\triangleq\max_{t=1}^T \Delta_{m,t}$.
%which is measured by the maximum difference of the observation values from any two adjacent datasets $\mathcal{D}_m$ and $\mathcal{D}_m^\prime$ with other variables fixed. In this section, the outputs of BS's $m$-th information extractor in \eqref{eq13} w.r.t. the input dataset $\mathcal{D}_m$ and $\mathcal{D}_m^\prime$ are denoted by $\rv_{m,t}(\mathcal{D}_m;\cdot)$ and $ \rv_{m,t}(\mathcal{D}_m^\prime;\cdot)$, respectively. The sensitivity is given by

%\lemma{For any given $\delta_m\in[0,1]$, the $m$-th WD is $(\e_m^{\text{WD}},\delta_m)$-DP if
%\begin{align}\label{eq24}
% \sqrt{\sum_{m^\prime=1}^M |\fv_m^H\hv_{m^\prime}|^2|s_{m^\prime,2}|^2+\sigma_z^2}\geq\frac{\Delta_m\sqrt{2T\ln(1/\delta_m)}}{\epsilon_m^{\text{WD}}}.
%\end{align}
%}
%\begin{IEEEproof}
%	The result is obtained from \cite[Theorem 1]{TMC_DP} by noticing that the power of effective noise in $\rv_{m,t}$ equals to $\sum_{m^\prime=1}^M |\fv_m^H\hv_{m^\prime}|^2|s_{m^\prime,2}|^2+\sigma_z^2$; see \eqref{eq13}.
%\end{IEEEproof}	
By bounding the maximum sensitivity and applying the result in \cite[Theorem 1]{TMC_DP}, we obtain the analytical expression of $\e_m^{(\text{BS})}(\cdot)$ in the following result.
%reach the following sufficient condition for \eqref{P1b}:
%
\corollary{\label{them2}Suppose that the BS applies any given information extractor $\fv_m$.  
	For any $\delta_m\in[0,1]$ fixed, the $m$-th WD can achieve $(\e_m^{(\text{BS})}(\cdot),\delta_m)$-DP at the BS after $T$ training rounds for any $\e_m^{(\text{BS})}(\cdot)$ satisfying
	\begin{align}\label{eq27}
		(\e_m^{(\text{BS})}(\cdot)) ^2\geq\frac{8|\fv_m^H\hv_m|^2|s_{m,1}|^2dT\ln(1/\delta_m)}{ K_m^2\left( {\sum_{m^\prime=1}^M |\fv_m^H\hv_{m^\prime}|^2|s_{m^\prime,2}|^2+\sigma_z^2}\right) }.
	\end{align}
}
\begin{IEEEproof}
	We first bound the maximum sensitivity $\Delta_m$ as \eqref{eq25}, shown on top of this page.
	\begin{figure*}
		
		\begin{align}\label{eq25}
			&\Delta_m=\max_t\max_{\mathcal{D}_m,\mathcal{D}'_m}\big\|\fv_m^H\hv_m\frac{s_{m,1}}{L}\mathbf{g}_{m,t}(\mathcal{D}_m;\cdot)-\fv_m^H\hv_m\frac{s_{m,1}}{L}\mathbf{g}_{m,t}(\mathcal{D}'_m;\cdot)\big\|_2\nonumber\\
			&=\frac{|\fv_m^H\hv_m||s_{m,1}|}{L}\max_t\max_{\substack{(\uv_{m,k},v_{m,k})\in\mathcal{D}_m\\(\uv'_{m,k},v'_{m,k})\in\mathcal{D}'_m}}\|\frac{1}{K_m}(\nabla f(\wv_t;\uv_{m,k},v_{m,k})-\nabla f(\wv_t;\uv'_{m,k},v'_{m,k}))\|_2\nonumber\\
			%		&\overset{(a)}\leq\frac{|\fv_m^H\hv_m||s_{m,1}|}{LK_m} \max_t\max_{\substack{(\uv_{m,k},v_{m,k})\in\mathcal{D}_m\\(\uv'_{m,k},v'_{m,k})\in\mathcal{D}'_m}}\left( \|\nabla f(\wv_t;\uv_{m,k},v_{m,k})\|_2+\|\nabla f(\wv_t;\uv'_{m,k},v'_{m,k})\|_2\right) \nonumber\\
			&\overset{(a)}\leq\frac{|\fv_m^H\hv_m||s_{m,1}|}{LK_m} \max_t\left( \max_{(\uv_{m,k},v_{m,k})}\|\nabla f(\wv_t;\uv_{m,k},v_{m,k})\|_2+\max_{(\uv'_{m,k},v'_{m,k})}\|\nabla f(\wv_t;\uv'_{m,k},v'_{m,k})\|_2\right)
			\overset{(b)}\leq \frac{2\sqrt d|\fv_m^H\hv_m||s_{m,1}|}{K_m}.
		\end{align}
		\hrulefill
	\end{figure*}
	where $(a)$ follows from the triangle inequality, and $(b)$ follows from Assumption \ref{assump1}.
	{Furthermore, the result in \cite[Theorem 1]{TMC_DP} bounds the cumulative privacy loss by moments accountant}.
	Specifically, the $m$-th WD is $(\e_m,\delta_m)$-DP with
	\begin{align}\label{eq24}
		{\sum_{m^\prime=1}^M |\fv_m^H\hv_{m^\prime}|^2|s_{m^\prime,2}|^2+\sigma_z^2}\geq\frac{\Delta_m\sqrt{2T\ln(1/\delta_m)}}{\epsilon_m},
	\end{align}
	where the left-hand side is the variance of the Gaussian noise $q_{t}[i]$.
	%	{\color{red} How to model the different bound?}
	\eqref{eq27} follows by combining \eqref{eq25} and  \eqref{eq24}.
\end{IEEEproof}

Corollary \ref{them2} reveals the impact of communication design on privacy loss.  First, the accumulated privacy loss of WD $m$ increases when the transmit power for local gradients (i.e., $s_{m,1}$) increases, or the artificial noise variances (i.e., $\{s_{m,2}\}_m$) decrease.
%Theorems \ref{theorem1} and \ref{them2} quantify  the critical effect of  transmit power allocation between gradient signals and artificial noise in determining the training loss  and the privacy loss. This choice reflects the fundamental learning-privacy trade-off in the FL system. 
Second, the denominator of \eqref{eq27} shows that the minimum privacy loss of WD $m$ is influenced by the receive beamforming gains w.r.t. the channels of all the WDs, \ie $|\fv_m^H\hv_{m^\prime}|^2,\forall m^\prime$. This is because artificial noises from different WDs are superposed at the receiver in over-the-air model uploading. Corollary \ref{them2} provides a tractable privacy loss bound that we shall use to simplify the FL system optimization problem.

%important properties of the privacy level of the FL system. First, the transmitting power split reflected by 

%{\color{red}Shall we provide some insights regarding Theorem 2 (e.g., the impact of $T$ and $s_{m,1},s_{m,2}$ on the accumulated privacy loss) here? We can also mention here that the lower bound of $\e_m^{(\text{BS})}(\cdot)$ in \eqref{eq27} is considered in the following. }

\subsection{Problem Simplification}
With the analytical results in Theorem \ref{theorem1} and Corollary \ref{them2}, we are ready to simplify the original problem (P1).  First, we replace the original objective $\E[F(\wv_T)]$ with its upper bound on the r.h.s. of \eqref{upperbound}. Then, by leveraging the closed-form expression of $\epsilon_m^{(\text{BS})}(\fv_m,\{s_{m,1}, s_{m,2}\}_{m=1}^M)$ in Theorem 2, the problem in \eqref{eq14} for the optimal private information extraction is given by
%the sufficient DP  condition in \eqref{eq27} is used to replace the DP constraint  \eqref{P1b} with $\e_m=\e_m^{(\text{WD})}$ and $\fv_m=\fv_m^\star$. Finally, to simplify the information extraction optimization in \eqref{eq14}, we approximate the objective in \eqref{eq14} by the r.h.s. of  \eqref{eq27}, as 
\begin{align}\label{eq26}
	\fv^\star_m
	%	\left(\{s_{m,1},s_{m,2}\}_{m=1}^M\right)
	=\argmax_{\norm{\fv_m}_2^2=1}
	\frac{8|(\fv_m)^H\hv_m|^2|s_{m,1}|^2dT\ln(1/\delta_m)}{ K_m^2\left( {\sum_{m^\prime=1}^M |(\fv_m)^H\hv_{m^\prime}|^2|s_{m^\prime,2}|^2+\sigma_z^2}\right) }, \forall m.
\end{align}

To summarize, the simplified system optimization problem is given as follows.\footnote{{The optimization in (P2) requires the knowledge of the Lipschitz constant $\omega$.  In practice, $\omega$ can either be directly calculated from the loss function and the training data or estimated via its statistics. Alternatively, it can be determined through trial and validation.}}
\begin{subequations}
	\begin{align}
		&(\text{P}2):	\min_{\eta,\fv_0,\{s_{m,1},s_{m,2}\}_{m=1}^{M}}A(\{s_{m,2}\},\eta,\fv_0)\frac{1-B^{T}}{1-B}\nonumber\\
		&~~~~~~~~~~~~~~~~~~+\sum_{t=1}^TB^{T-t}{C_{t}}(\{s_{m,1}\},\eta,\fv_0),\label{P2a}\\
		&\text{s.t. } 
		{\sum_{m^\prime=1}^M |(\fv_m^\star)^H\hv_{m^\prime}|^2|s_{m^\prime,2}|^2+\sigma_z^2}	\nonumber\\
		&~~~~~~~~\geq\frac{8|(\fv^\star_m)^H\hv_m|^2|s_{m,1}|^2dT\ln(1/\delta_m)}{ K_m^2( \epsilon_m^{(\text{WD})}) ^2}, \forall m, \label{P2b}\\
		&~~~~\eqref{P1d}-\eqref{P1e}.
	\end{align}
\end{subequations}
Here, we have dropped the constant terms in the objective. {(P2) inherently possesses feasibility given the presence of trivial feasible points, specifically when $s_{m,1}=s_{m,2}=0,\forall m$.} {In contrast to the original problem (P1), (P2) targets the minimization of an upper bound of the original objective while being confined to a constrained feasible subset, thus serving as a surrogate for (P1).} Despite simplification through Theorem \ref{theorem1} and Corollary \ref{them2}, (P2) is still difficult to solve due to the non-convex objective function and constraints, which is further exacerbated by the $M$ nested non-convex sub-problems for the optimization of the private information extractors $\{\fv_m^\star\}_{m=1}^M$ in \eqref{P2b}.
%We note that the original constraint \eqref{P1b} is automatically fulfilled by \eqref{P2b} and \eqref{eq26}. 

Before studying the general solution to (P2), we first consider a special case with a single-antenna BS, \ie $N=1$, in Section \ref{sec4} to shed light on the general case.  Although (P2) is still non-convex with $N=1$, we show in Section \ref{sec4} that there always exists a closed-form optimal solution.

\section{Special Case: Transceiver Optimization With A Singe-Antenna BS}\label{sec4}
We present the solution to (P2) when $N=1$ in this section. 
With $N=1$, the vectors $\fv_0,\fv_m,\hv_m,\forall m,$ regress to scalars $f_0,f_m,h_m$, respectively. For any $\eta$ and $\{s_{m,1},s_{m,2}\}_{m=1}^M$, it can be verified that any $f_0$ with  $|f_0|^2=1$ is optimal to (P2), and any $f_m^\star$ with $|f_m^\star|^2=1,\forall m,$ is optimal to \eqref{eq26}.
Without loss of generality, we set $f_0=f_m=1,\forall m$. 
The problem (P2) can be rewritten as
\begin{subequations}\label{eq61}
	\begin{align}
		&(\text{P2a}): 		\min_{\eta,\{s_{m,1},s_{m,2}\}_{m=1}^{M}}A(\{s_{m,2}\},\eta,1)\frac{1-B^{T}}{1-B}\nonumber\\
		&~~~~~~~~~~~~~~~~+\sum_{t=1}^TB^{T-t}{C_{t}}(\{s_{m,1}\},\eta,1),\label{eq61a}\\
		\text{s.t. } 
		&  {\sum_{m^\prime=1}^M |h_{m^\prime}|^2|s_{m^\prime,2}|^2+\sigma_z^2}	\geq\frac{|h_m|^2|s_{m,1}|^2T\varphi_m}{ K_m^2}, \forall m,\\
		& \eta>0,|s_{m,1}|^2+|s_{m,2}|^2\leq P_{\text{max}},\forall m,
	\end{align}
\end{subequations}
where the auxiliary variables are simplified as
%\begin{align}
$A(\{s_{m,2}\},\eta,1)={\left(\sum_{m=1}^M|h_m|^2|s_{m,2}|^2+\sigma_z^2\right)}d/{2\omega K^2\eta}$, 
$C_{t}(\{s_{m,1}\},\eta,1)=\frac{1}{2\omega K^2}\sum_{i=1}^d\mathbb{E}[|\sum_{m=1}^M(K_m-\frac{h_ms_{m,1}}{\sqrt{\eta}L})g_{m,t}[i]|^2]$, and 
$	\varphi_m\triangleq\frac{8d\ln(1/\delta_m)}{(\epsilon_m^{(\text{WD})})^2}, \forall m$.
%\end{align}
%%
%\begin{subequations}
%	\begin{align}
%		(\text{P}2a):	\min_{\eta,\{s_{m,1},s_{m,2}\}_{m=1}^{M}}&\sum_{t=1}^TB^{T-t}{C_{t}}(\{s_{m,1}\},\eta,1),\\
%		\text{s.t. } 
%		&  {\sum_{m^\prime=1}^M |h_{m^\prime}|^2|s_{m^\prime,2}|^2+\sigma_z^2}	\geq\frac{8|h_m|^2|s_{m,1}|^2dT\ln(1/\delta_m)}{ K_m^2\epsilon_m^2}, \forall m, \label{P3b}\\
%		& |s_{m,1}|^2+|s_{m,2}|^2\leq P_{\text{max}},\forall m,\\
%		& \eta>0, T\geq 0, T\in\mathcal{Z}.
%	\end{align}
%\end{subequations}
%
%We further define a constant parameter $T_0\triangleq \frac{\sigma_z^2}{P_{\text{max}}\max_{m}\varphi_m}\max_m\frac{K_m^2}{|h_m|^2}$.
Note that $\varphi_m$ is a constant depending on the DP requirement of WD $m$. Define $T_0\triangleq \frac{\sigma_z^2}{P_{\text{max}}\max_{m}\varphi_m}\max_m\frac{K_m^2}{|h_m|^2}$. The following proposition presents the optimal condition to (P2a).
\proposition[Sufficient and necessary optimality condition]{\label{pro1}Given $T\geq 0$, a solution $\{\eta,\{s_{m,1},s_{m,2}\}_m\}$ is optimal to (P2a) \emph{if and only if} the following conditions hold:
	\begin{itemize}
		\item When  $T\geq T_0$,
		\begin{align}
			%\left\{
			%\begin{aligned}
			&\eta=\frac{\sum_{m^\prime=1}^M|h_{m^\prime}|^2|s_{m^\prime,2}|^2+\sigma_z^2}{L^2T\max_{m}\varphi_m},\\
			% \end{align}
			%\begin{align}
			&\frac{|h_m|^2|s_{m,1}|^2}{K_m^2}=\frac{\sum_{m^\prime=1}^M|h_{m^\prime}|^2|s_{m^\prime,2}|^2+\sigma_z^2}{T\max_{m}\varphi_m},\forall m,\\
			&	\frac{\sum_{m^\prime=1}^M|h_{m^\prime}|^2|s_{m^\prime,2}|^2+\sigma_z^2}{T\max_{m}\varphi_m}\leq \min_m\frac{|h_m|^2(P_{\text{max}}-|s_{m,2}|^2)}{K_m^2}.\label{case1}
			%\end{aligned}
			%\right.
		\end{align}
		\item When $T< T_0$,
		\begin{align}
			%\left\{
			%\begin{aligned}
			&	\eta=\frac{P_{\text{max}}}{L^2}\min_m\frac{|h_m|^2}{K_m^2},\label{case2a}\\	
			&	s_{m,1}=\frac{\sqrt\eta LK_m}{|h_m|^2}\bar h_m,\forall m,\\
			&	s_{m,2}=0, \forall m,\label{case2}
			%\end{aligned}
			%\right.
		\end{align}
		where $\bar h_m$ is the conjugate of $h_m$.
	\end{itemize}
}
\begin{IEEEproof}
	See Appendix \ref{appb}.
\end{IEEEproof}

Proposition \ref{pro1} indicates that there is more than one global optimum when $T\geq T_0$ as the equation system in \eqref{case1} is underdetermined. We note that 
one of the solutions to \eqref{case1} corresponds to $s_{m,2}=0,\forall m$ from the definition of $T_0$. Together with the case of $T< T_0$, we obtain the following result from Proposition \ref{pro1}. 
\corollary[Zero-artificial-noise property]{
	With a single-antenna BS, an optimal solution to (P2a) is given by
	%This observation does not hold for $N>1$.
	%\subsection{Closed-Form Optimal Solution}
	%For any $T$, we choose one set of solutions satisfying the conditions in Proposition \ref{pro1} as 
	\begin{subequations}\label{optimal_exceptT}
		\begin{align}
			&	\eta^*=	\left\{
			\begin{aligned}
				&\frac{\sigma_z^2}{L^2T\max_{m}\varphi_m},& \text{ if } T\geq T_0,\\
				&\frac{P_{\text{max}}}{L^2}\min_m\frac{|h_m|^2}{K_m^2},& \text{ otherwise},
			\end{aligned}
			\right.\\
			&	s_{m,1}^*=\frac{\sqrt\eta LK_m}{|h_m|^2}\bar h_m,\forall m,\\
			&	s_{m,2}^*=0, \forall m.
		\end{align}
	\end{subequations}
}

The result in Lemma 1 reveals an interesting observation: The WDs do not need to add extra artificial noise to preserve DP when $N=1$. Instead, they can achieve any privacy level by \emph{coherently decreasing the transmit power $\{s_{m,1}\}$ of all the WDs}. This result coincides with the existing work on single-antenna receiver design in \cite{DPFL_DLiu}. 
We note that this property is helpful in practical implementation as it avoids extra hardware costs and power consumption in introducing additional DP mechanisms.  
\section{Transceiver Optimization With A Multiple-Antenna BS}
\label{sec5}
Section \ref{sec4} verifies the zero-artifical-noise property for $N=1$. However, as we shall demonstrate, this favorable property does not always hold for general multi-antenna cases.  Moreover, the BS-side information extractor design problem in \eqref{eq26} becomes non-trivial with $N>1$. 
In this section, we derive a closed-form optimal solution to \eqref{eq26}, followed by a sub-optimal solution to Problem (P2) based on alternating optimization.
%Here, we demonstrate that zero artificial noise may not be optimal by the following example. 
%\example{123}

\subsection{Optimal BS-Side Private Information Extractors}
\label{sec_5a}

For given $\{s_{m,1},s_{m,2}\}_m$ and $N\geq 1$, the problem in \eqref{eq26} can be recast as
\begin{align}
	\fv^\star_m&= \argmax_{\norm{\fv_m}_2^2=1} \frac{8|\fv_m^H\hv_m|^2|s_{m,1}|^2dT\ln(1/\delta_m)}{ K_m^2\left( {\sum_{m^\prime=1}^M |\fv_m^H\hv_{m^\prime}|^2|s_{m^\prime,2}|^2+\sigma_z^2}\right) }\nonumber\\
	%	&=\argmin_{\norm{\fv_m}_2^2=1}\frac{  {\sum_{m^\prime=1}^M |\fv_m^H\hv_{m^\prime}|^2|s_{m^\prime,2}|^2+\sigma_z^2} }{|\fv_m^H\hv_m|^2|s_{m,1}|^2}\nonumber\\
	&=\argmin_{\norm{\fv_m}_2^2=1}\frac{  {\sum_{m^\prime\neq m} |\fv_m^H\hv_{m^\prime}|^2|s_{m^\prime,2}|^2+\sigma_z^2} }{|\fv_m^H\hv_m|^2|s_{m,1}|^2}+\frac{|s_{m,2}|^2}{|s_{m,1}|^2}\nonumber\\
	&=\argmax_{\norm{\fv_m}_2^2=1}\frac{|\fv_m^H(\hv_ms_{m,1})|^2}{  {\sum_{m^\prime\neq m} |\fv_m^H(\hv_{m^\prime}s_{m^\prime,2})|^2+\sigma_z^2} }.\label{eq41}
\end{align}
Define $\widetilde\Hv_m\triangleq[\hv_1s_{1,2},\cdots,\hv_{m-1}s_{m-1,2},\hv_ms_{m,1},\hv_{m+1}s_{m+1,2},$ $\cdots,\hv_Ms_{M,2}]\in\Complex^{N\times M}$ for $\forall m$. The objective in \eqref{eq41} can be seen as the signal-to-interference-plus-noise (SINR)  ratio of an $M$-user uplink communication system under channel matrix $\widetilde\Hv_m$ and $M$ unit-modular Gaussian transmit signals, where $\fv_m^H$ is a linear receive estimator for the $m$-th signal.
It has been shown in \cite{MIMO_DTse,Est_Theory} that the minimum mean-square-error (MMSE) estimator  maximizes the SINR of such  a system, \ie
\begin{align}\label{temp42}
	\fv^\star_m=\frac{1}{c_m}\left[\widetilde\Hv_m\left( \widetilde\Hv_m^H\widetilde\Hv_m+\sigma^2_z\Iv_M\right)^{-1} \right]_{:,m},\forall m,
\end{align}
where $[\Xv]_{:,m}$ is the $m$-th column of matrix $\Xv$, and $c_m$ is a normalization factor to ensure $\norm{\fv^\star_m}_2=1$.

{As shown in \eqref{temp42}, the MIMO information extractor effectively isolates the signal of a targeted WD from the interference caused by other WDs. This is fundamentally different from the MISO-based model aggregation, where the received signals from different WDs invariably merge.}

\subsection{Zero Artificial Noise Is Generally Sub-Optimal for $N> 1$}
We argue in this section that zero artificial noise (i.e., $s_{m,2}=0,\forall m$) may not be optimal to (P2) with $N>1$. We demonstrate this by using the following example.
\example{Let $M=2$, $N>1$, and the signal-to-noise ratio (SNR) be high (\ie $\sigma_z^2\to 0$). We consider a scenario where the channels of WDs $1$ and $2$ are orthogonal, \ie $\hv_1^H\hv_2=0$. In this case, it can be verified that the $\fv_m^\star$ in \eqref{temp42} is given by $\fv_m^\star=\hv_m/\norm{\hv_{m}}_2$. 
	%	Moreover, we assume that WD $1$ has a much more stringent DP requirement than WD $2$, \ie $\epsilon_1^{(\text{WD})}\ll \epsilon_2^{(\text{WD})}$.
	%	suffers from deep fading, \ie $\norm{\hv_2}_2\approx 0$ and $\norm{\hv_1}_2\gg \norm{\hv_2}_2$. 
	The constraint \eqref{P2b} becomes, for $m=1,2$,
	\begin{align}
		|s_{m,2}|^2	\geq\frac{8\norm{\hv_{m}}_2^2dT\ln(1/\delta_m)}{ K_m^2( \epsilon_m^{(\text{WD})}) ^2} {|s_{m,1}|^2}. 
	\end{align}
	In this case, a solution corresponding to $s_{m,2}=0$ leads to $s_{m,1}=0$. In other words, the two WDs will upload nothing to the BS. Clearly, this is not a good choice for FL model training.  Instead, one can fix $s_{m,1}\neq 0,\forall m,$ and adjust $s_{m,2}$ proportionally to  $ \frac{\ln(1/\delta_m)}{ K_m^2( \epsilon_m^{(\text{WD})}) ^2} $, which shall yield a better training performance.  
}

Example 1 provides an important intuition on why the zero-artificial-noise solution is not preferred for $N>1$.  Due to the employment of $M$ information extractors for $M$ datasets, the BS can {adjust $\fv_m^\star$ based on $\hv_m$} such that the inter-user interference (or equivalently, the artificial noise contributed from the other users) is minimized. In this case, WD $m$ has no choice but to add non-zero artificial noise to fulfill its own privacy requirement.
%hence non-zero artificial noise is preferred. 
Notice that such fundamental difference between the single-antenna and multi-antenna cases is due to the fact that the single-antenna BS cannot distinguish transmitted signals from different WDs with over-the-air model aggregation \cite{9174426}. 

\subsection{Proposed Algorithm to (P2) via  Alternating Optimization}
As we cannot directly extend the result in Corollary 2 
%\eqref{optimal_exceptT}
to the multi-antenna case, we propose an alternating optimization algorithm to sub-optimally solve (P2). The following proposition simplifies (P2).
\proposition[Necessary optimality condition]{\label{pro3}The optimal solution to Problem (P2) must satisfy $C_t(\{s_{m,1},\eta,\fv_0\})=0,\forall t$, with $C_t(\cdot)$ defined in \eqref{C}. Moreover, the variables $\{s_{m,1}\}$ are optimal only if 
	\begin{align}\label{temp43}
		s_{m,1}=\frac{\sqrt\eta LK_m}{|\fv_0^H\hv_m|^2}\overline{\fv_0^H\hv_m},\forall m.
	\end{align}
}
\begin{IEEEproof}
	The result can be obtained by following the arguments in \eqref{eqaapb1}--\eqref{eq66} of Appendix \ref{appb}.
\end{IEEEproof}

%	\subsubsection{Given $T$, optimizing $\fv_0$, $s_{m,2}$ and $\eta$}:
By applying Proposition \ref{pro3} and dropping the constant terms, (P2) is simplified to
\begin{align}\label{eq44}
	&\min_{\fv_0,\{s_{m,2}\},\eta}~~A(\{s_{m,2}\},\eta,\fv_0)=\frac{\left(\sum_{m=1}^M|\fv_0^H\hv_m|^2|s_{m,2}|^2+\sigma_z^2\right)d}{2\omega K^2\eta}\nonumber\\
	&\text{s.t. } \frac{|(\fv_m^\star)^H\hv_m|^2TL^2\varphi_m}{|\fv_0^H\hv_m|^2}\eta\leq \sum_{m^\prime=1}^M |(\fv_m^\star)^H\hv_{m^\prime}|^2|s_{m^\prime,2}|^2+\sigma_z^2,\forall m,\nonumber\\
	&~~~~|s_{m,2}|^2+\frac{K_m^2L^2}{|\fv_0^H\hv_m|^2}\eta\leq P_{\text{max}},\forall m,\nonumber\\
	&~~~~\eta>0,\norm{\fv_0}^2_2=1.
\end{align}
%\begin{itemize}
%	\item If $T\leq \frac{\sigma_z^2}{P_{\text{max}}}\frac{\min_m |\fv_0^H\hv_m|^2/(\varphi_m|\fv_m^H\hv_m|^2)}{\min_m |\fv_0^H\hv_m|^2/K_m^2}$, we have (conventional over-the-air computation)
%	\begin{align}\label{temp45}
%		&\eta=\frac{P_{\text{max}}}{L^2}\min_m\frac{|\fv_0^H\hv_m|^2}{K_m^2},\\
%		&s_{m,2}=0,\forall m,\\
%		&\fv_0=\min_{\norm{\fv_0}=1} \max \frac{K_m^2}{|\fv_0^H\hv_m|}.
%	\end{align}

We solve \eqref{eq44} in an alternating fashion as follows. On one hand, given $\{s_{m,2}\}$, we define $\tau\triangleq \frac{1}{\eta}$ and  $\Fv\triangleq \fv_0\fv_0^H/\eta$ stratifying $\Fv\succeq \bf 0$ and $\rank(\Fv)=1$. The problem in \eqref{eq44} can be transformed into a semi-definite programming (SDP) problem, as
\begin{subequations}
	
	\label{eqsub1}
	\begin{align}
		\min_{\Fv,\tau}&~~\sum_{m=1}^M|s_{m,2}|^2\tr\left( \Fv\hv_m\hv_m^H\right) +\sigma_z^2\tau\label{eqsub1a}\\
		\text{s.t. }& \frac{|(\fv_m^\star)^H\hv_m|^2TL^2\varphi_m}{\sum_{m^\prime=1}^M |(\fv_m^\star)^H\hv_{m^\prime}|^2|s_{m^\prime,2}|^2+\sigma_z^2}\leq \tr\left( \Fv\hv_m\hv_m^H\right),\forall m	\label{eqsub1b}\\
		&\frac{K_m^2L^2}{P_{\text{max}}-|s_{m,2}|^2}\leq \tr\left( \Fv\hv_m\hv_m^H\right),\forall m,	\label{eqsub1c}\\
		&\tr(\Fv)=\tau,	\label{eqsub1d}\tau>0, \Fv\succeq {\bf 0},\\
		&\rank(\Fv)=1.\label{eqsub1f}
	\end{align}
\end{subequations}
Note that the only non-convex constraint is the rank-one constraint in \eqref{eqsub1f}. We adopt the DC optimization approach to solve \eqref{eqsub1}. Specifically, we have $\rank(\Fv)=1\Leftrightarrow\tr(\Fv)-\norm{\Fv}_2=0$ \cite{FL_1}, where $\norm{\Fv}_2$ is the spectral norm of $\Fv$. Motivated by this, we penalize the term  $\tr(\Fv)-\norm{\Fv}_2$ in the objective function as
\begin{align}\label{eqsub2}
	\min_{\Fv,\tau}&~~\sum_{m=1}^M|s_{m,2}|^2\tr\left( \Fv\hv_m\hv_m^H\right) +\sigma_z^2\tau+\rho (\tr(\Fv)-\norm{\Fv}_2)\nonumber\\
	\text{s.t. }& \eqref{eqsub1b}-\eqref{eqsub1d},
\end{align}
where $\rho\geq0$ is the predefined penalty parameter. To solve \eqref{eqsub2}, we iteratively linearize $-\rho\norm{\Fv}$ via majorization-minimization. That is, for iteration $j=1,2,\cdots,J$, we approximate $-\rho\norm{\Fv}$ by a  surrogate function based on the current value $\Fv^{(j)}$, \ie
\begin{align}\label{eq42}
	&-\rho\norm{\Fv}\leq -\rho\norm{\Fv^{(j)}}+\tr(\Fv\cdot\partial_{\Fv^{(j)}}(-\rho\norm{\Fv}))\nonumber\\
	&=-\rho\norm{\Fv^{(j)}}-\rho\tr(\Fv\zetav^{(j)}(\zetav^{(j)})^H),
\end{align}
where $\partial_{\Fv^{(j)}}(-\rho\norm{\Fv})$ is the subgradient of $-\rho\norm{\Fv}$ at $\Fv^{(j)}$, and $\zetav^{(j)}$ is the principal eigenvector of $\Fv^{(j)}$. Leveraging \eqref{eq42}, we construct the following convex problems 
\begin{align}\label{eqsub3}
	(\Fv^{(j+1)},\tau^{(j+1)})&=\argmin_{\Fv\neq {\bf 0},\tau}\sum_{m=1}^M|s_{m,2}|^2\tr\left( \Fv\hv_m\hv_m^H\right) +\sigma_z^2\tau\nonumber\\
	&~~~~~~~~~~+\rho \tr\left( \Fv\left( \Iv_N-\zetav^{(j)}(\zetav^{(j)})^H\right) \right),\nonumber\\
	&~~~~	\text{s.t.}~\eqref{eqsub1b}-\eqref{eqsub1d},
\end{align}
which can be efficiently solved via off-the-shelf solvers, such as CVX \cite{cvx}. After obtaining $\Fv^{(J)}$, the solution to $\fv_0$ is obtained as its principal eigenvector by the Cholesky or eigenvalue decomposition of $\Fv^{(J)}$.

On the other hand, given $\fv_0$ and $\eta$, the problem of optimizing   $\{s_{m,2}\}$ is given by
\begin{align}\label{eq46}
	&\min_{\{s_{m,2}\}_{\forall m}}~\sum_{m=1}^M|\fv_0^H\hv_m|^2|s_{m,2}|^2\nonumber\\
	&\text{s.t. } \frac{|(\fv_m^\star)^H\hv_m|^2TL^2\varphi_m}{|\fv_0^H\hv_m|^2\eta}\!-\!\sigma_z^2\!\leq\! \sum_{m^\prime=1}^M |(\fv_m^\star)^H\hv_{m^\prime}|^2|s_{m^\prime,2}|^2,\forall m,\nonumber\\
	&~~~~0\leq |s_{m,2}|^2\leq P_{\text{max}}-\frac{K_m^2L^2}{|\fv_0^H\hv_m|^2}\eta,\forall m.
\end{align}
Problem \eqref{eq46} is a linear programming problem, which can be efficiently solved, \eg by the interior point method \cite{LP}.

We summarize the overall algorithm in Algorithm \ref{alg1}. {We show in Appendix \ref{appc} that Algorithm \ref{alg1} makes the objective in \eqref{eq44} converge to a stationary value for some $\rho\geq 0$ in \eqref{eqsub2}.}
To accelerate the convergence, in Step 11 an early-stopping criterion is added to save computational time. {In implementing Algorithm \ref{alg1}, the BS optimizes $\eta$, $\fv_0$ and $\{s_{m,1},s_{m,2}\}_m$ based on the CSI on $\{\hv_m\}$. Once the optimization converges, the BS transmits $\{s_{m,1},s_{m,2}\}_m$ to each WD $m$ via a reliable link for over-the-air model uploading. This procedure eliminates the need for CSI feedback to the WDs.
}

\begin{algorithm}[!t]
	\caption{The proposed algorithm to (P2).}
	\label{alg1}
	%	\label{algsca}
	\begin{algorithmic}[1]
		\STATE\textbf{Input:}  
		$T$, $\{\e^{\text{(WD)}}_m,\delta_m,\hv_m,K_m\}_{\forall m}$, $P_{\text{max}},\sigma_z^2,\mu,\omega,L,\rho$.
		\STATE Initialize  $\fv_0$ and $\{s_{m,1},s_{m,2}\}.$
		%		=\sqrt{\frac{P_{\text{max}}}{2}}, 
		%		$\forall m$.
		\STATE \textbf{for} $\mbox{iter}=1,2,\cdots,I$\\
		\STATE~~Update $\fv_m^\star$ by \eqref{temp42};\\
		\STATE ~~\textbf{for} $j=1,2,\cdots,J$\\
		\STATE ~~~~	Update $\Fv^{(j)}$ and $\tau$ by solving \eqref{eqsub3};
		\STATE~~\textbf{end for} \\
		\STATE~~Compute $\eta=1/\tau$ and $\fv_0$ as the principal eigenvector of $\Fv^{(J)}$;\\
		\STATE~~Update $\{s_{m,2}\}$ by solving \eqref{eq46}; \\
		\STATE~~Update $\{s_{m,1}\}$ by \eqref{temp43};\\
		%						\STATE~~Update $T$ by \eqref{eq51};\\
		\STATE~~\textbf{if} {$|A(\cdot;\mbox{iter})-A(\cdot;\mbox{iter}-1)|\leq 10^{-4}$}, \textbf{early stop};\\
		\STATE\textbf{end for} \\
		\STATE	\textbf{Output:} {$\eta, \fv_0$, and $\{s_{m,1},s_{m,2},\fv_m^\star\}_{\forall m}$.}
	\end{algorithmic}
\end{algorithm}
\section{Numerical Results}\label{sec6}
In this section, we present numerical experiments to evaluate the performance of the proposed FL design.

\subsection{Simulation Setup}
We simulate a ridge regression problem using a synthetic dataset with $K=1000$ data samples and $d=20$ model parameters. The loss function is given by
\begin{align}\label{eq45}
	F(\wv)=\frac{1}{2K}\sum_{k=1}^K \norm{v_k-\wv^T\uv_k}_2^2+ \frac{\varphi}{2} \norm{\wv}_2^2,
\end{align}
where $\uv_k\in\Real^{d\times 1}$ and $v_k\in\Real$ are the input and output of the $k$-th data sample, respectively, and $\varphi=10^{-3}$ is the regularization parameter. 
%We generate $K=1000$ data samples as follows. 
The input vector $\uv_k$ is i.i.d. drawn from $\Norm({\bf 0},\Iv_{d})$. The output $v_k$ is generated as $v_k=\wv_{\text{true}}^T\uv_k+n_k$, where $\wv_{\text{true}}\in\Real^{d\times 1}$ is the true model parameter vector randomly generated by following $\Norm({\bf 0},\Iv_{d})$, and $n_k\sim \Norm(0,0.2)$ represents the measurement error in data collection. The loss function in \eqref{eq45} satisfies Assumptions 2-4, where the hyper-parameters $\mu$ and $\omega$ are given by the smallest and largest eigenvalues of  $\frac{1}{K}\Uv^T\Uv+\varphi\Iv_{d}$, respectively, and $\Uv\triangleq[\uv_1,\cdots,\uv_K]^T$. Assumption \ref{assump1} is satisfied by the gradient clipping method in \cite{TMC_DP} with $L=0.1$. 
Moreover, the optimal solution to \eqref{eq45} is $\wv^\star=(\Uv^T\Uv+2K\varphi\Iv_d)^{-1}\Uv^T\vv$ with $\vv\triangleq[v_1,\cdots,v_K]^T$.  The training model $\wv$ is initialized as $\wv_0=\bf 0$.
Besides, we also study a real-world image classification task using the Fashion-MNIST \cite{Xiao2017} and CIFAR-10 \cite{Cifar} datasets and convolutional neural networks (CNNs) as detailed in Section \ref{sec_6d}. 

Unless otherwise specified, the simulation parameters are set as follows. The data samples are uniformly divided into $M=10$ WDs with equal data size $K_m=100$. We consider a  Raleigh fading channel model, where
the channel coefficient vectors $\hv_m$ are drawn from $\CN({\bf 0},\Iv_N)$.  The maximum transmit power $P_{\text{max}}$ is $1$ W,  and the AWGN noise $\sigma_z^2$ is determined by the SNR defined as $\text{SNR}\triangleq P_{\text{max}}/\sigma_z^2$. The number of FL training rounds is $T=30$, and the number of receive antennas at the BS is set to $N=20$. For the DP requirement at the WDs, $\delta_m$ is set to $10^{-3}$, and $\epsilon_m^{\text{(WD)}}$ equals to $\epsilon,\forall m,$ for some $\e$ specified later. 
All the results are averaged over $500$ Monte Carlo trials. For Algorithm \ref{alg1}, we set $\mbox{iter}=10$, $\rho=1$, $J=50$ and initialize $s_{m,1}$ uniformly in $[0,1],s_{m,2}=\sqrt{P_{\max}-|s_{m,1}|^2},$ and $\fv_0=\hv_1/\norm{\hv_1}_2$.  
%$15 dB,N=20$
\subsection{Impact of BS Information Extractor Design}
\begin{figure}[!t]
	\centering
	\includegraphics[width=2.8 in]{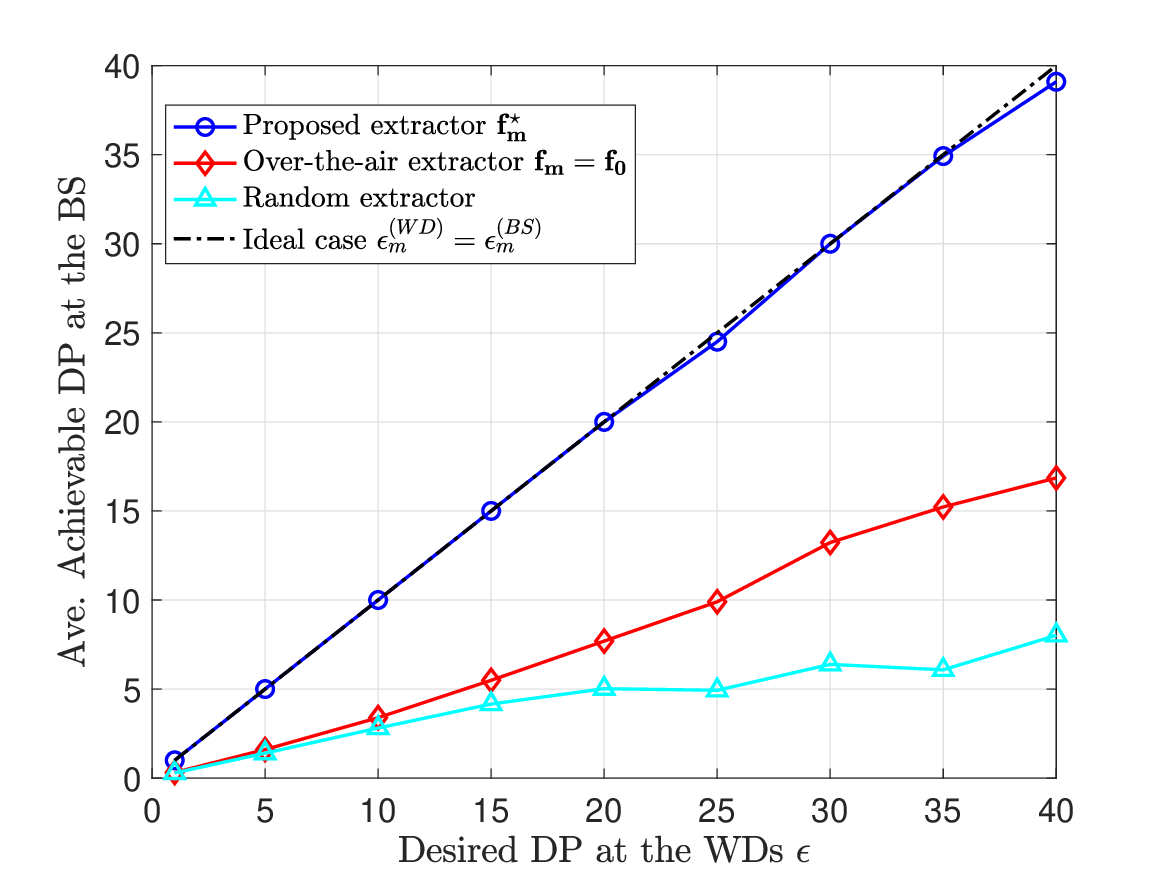}
	\caption{The average achievable DP at the BS, \ie $\frac{1}{M}\sum_m \e_m^{(\text{BS})}$, under different information extraction mechanisms. The transmission designing variables $\{s_{m,1},s_{m,2}\}$, $\eta$, and $\fv_0$ are optimized by Algorithm \ref{alg1}.}
	\label{fig2}
\end{figure}
As shown in Section \ref{sec_2b}, the BS-side information extraction mechanism design, \ie the optimization of  $\{\fv_m\}$ in \eqref{eq13}, is critical in MIMO-based FL systems. Specifically, by leveraging the separate private information extractors, the BS can beamform each combining vector towards the channel of a particular WD in order to reveal more sensitive information about its local dataset; see Section \ref{sec_5a}.
%{{This is different from the single-antenna receivers as the latter cannot distinguish the transmit signal of a particular WD from others owing to the superposition nature of over-the-air model aggregation \cite{9174426}.}} 
In this section, we demonstrate the impact of the BS-side information extraction process with different receive combining techniques. 

Fig. \ref{fig2} plots the relationship between the \emph{desired} and \emph{achieved} privacy levels. Here, we set the desired privacy level to $\e_m^{(\text{WD})}=\e,\forall m,$ for $\e\in[1,60]$, and examine the average achievable privacy at the BS as $\frac{1}{M}\sum_{m=1}^M \e_m^{(\text{BS})}$.\footnote{The exact value of the achievable privacy level $\e_m^{(\text{BS})}$ is not calculable. We approximate it with the bound in \eqref{eq27}.} As discussed in \eqref{eq14}, the BS uses $\{\fv_m\}$ to maximize $\e_m^{(\text{BS})}$ in the information extraction. Ideally, under the feasibility constraint \eqref{P1c}, the \emph{optimal} information extractors  should perfectly achieve the desired privacy level  as $\e_m^{(\text{BS})}=\e_m^{(\text{WD})},\forall m$. However, finding this ideal solution is computationally infeasible due to the intractability of \eqref{eq14}. In Fig. \ref{fig2}, we examine three solutions to the information extractors: a) the proposed solution $\{\fv_m^\star\}$ in \eqref{temp42}; b) the conventional over-the-air receive combining vector for model aggregation, i.e., $\fv_m=\fv_0,\forall m,$ with $\fv_0$ optimized by Algorithm \ref{alg1}; and c) a random information extractor $\fv_m$ drawn from $\Norm({\bf 0},\frac{1}{N}\Iv)$. We see that the achievable DP of the proposed closed-form solution is close to that of the ideal extractor. In contrast, the achievable  $\e_m^{(\text{BS})}$ of the conventional over-the-air receiver is much lower.
This highlights the necessity of the dedicated design in \eqref{temp42} for the information extraction mechanism instead of directly inferring private information from the estimated gradients after over-the-air model aggregation.

As a final remark, we emphasize that the BS-side information extraction mechanism has a critical impact on the WD-side DP mechanism design.  
%The BS is capable of employing multiple extractors with multiple receive antennas.
Taking the BS-side private information extractors into account, the WDs should impose larger artificial noise for privacy preservation in FL systems over  MIMO channels compared with the counterpart over MISO channels, as the latter cannot distinguish the transmit signal of a particular WD from others owing to the superposition nature of over-the-air model aggregation \cite{9174426}. %{{Otherwise, the desired DP cannot be achieved at the BS. }} 
{ On the other hand, the MIMO-based over-the-air receiver can harness high array gains to effectively minimize the aggregation error in local gradients. This implies that, although a MIMO system demands a higher degree of artificial noise at the WDs, its proficiency in enhancing the accuracy of over-the-air model aggregation may result in a larger learning convergence rate, as will be further illustrated in the following section.}
\subsection{Experiments on Synthetic Data}
We compare the FL training performance of the following schemes on the ridge regression model:
\begin{itemize}
	\item FL with a multiple-antenna BS and DP: The WDs impose DP requirements specified by  the value of $\e_m^{(\text{WD})}$. We optimize the transceiver design by Algorithm  1. 
	\item FL with a single-antenna BS and DP: Let $N=1$. We design the transceivers by the optimal solution in \eqref{optimal_exceptT}. Note that this solution aligns with the state-of-the-art result in \cite{DPFL_DLiu}.
	\item FL with a multiple-antenna BS but \emph{without} DP: Suppose there is no privacy constraint at the WDs, \ie $\e_m^{(\text{WD})}=\infty,\forall m$. We optimize $\{s_{m,1}\},\eta,$ and $\fv_0$ by Algorithm \ref{alg1} by ignoring the constraint \eqref{P1c}.  The solution is identical to that in the existing work on DP-unaware FL \cite{FL_2,FL_1}.  
	\item FL with a single-antenna BS but \emph{without} DP: Let $\e_m^{(\text{WD})}=\infty,\forall m,$ and $N=1$. The solution is given by \eqref{optimal_exceptT} with $T_0=\infty$. 
\end{itemize} 
The performance of these schemes is compared in terms of the \emph{normalized optimality gap} of the loss function \cite{DPFL_DLiu,DPFL_RIS} defined as $(F(\wv_T)-F(\wv^\star))/F(\wv^\star)$, with a focus on the trade-off between privacy enhancement and training performance.
For the DP-aware schemes, we set $\e_m^{(\text{WD})}=\e,\forall m,$ with $\e$ representing the privacy level.

\begin{figure}[!t]
	%\begin{minipage}[t]{0.49\linewidth}
	\centering
	\includegraphics[width=2.8 in]{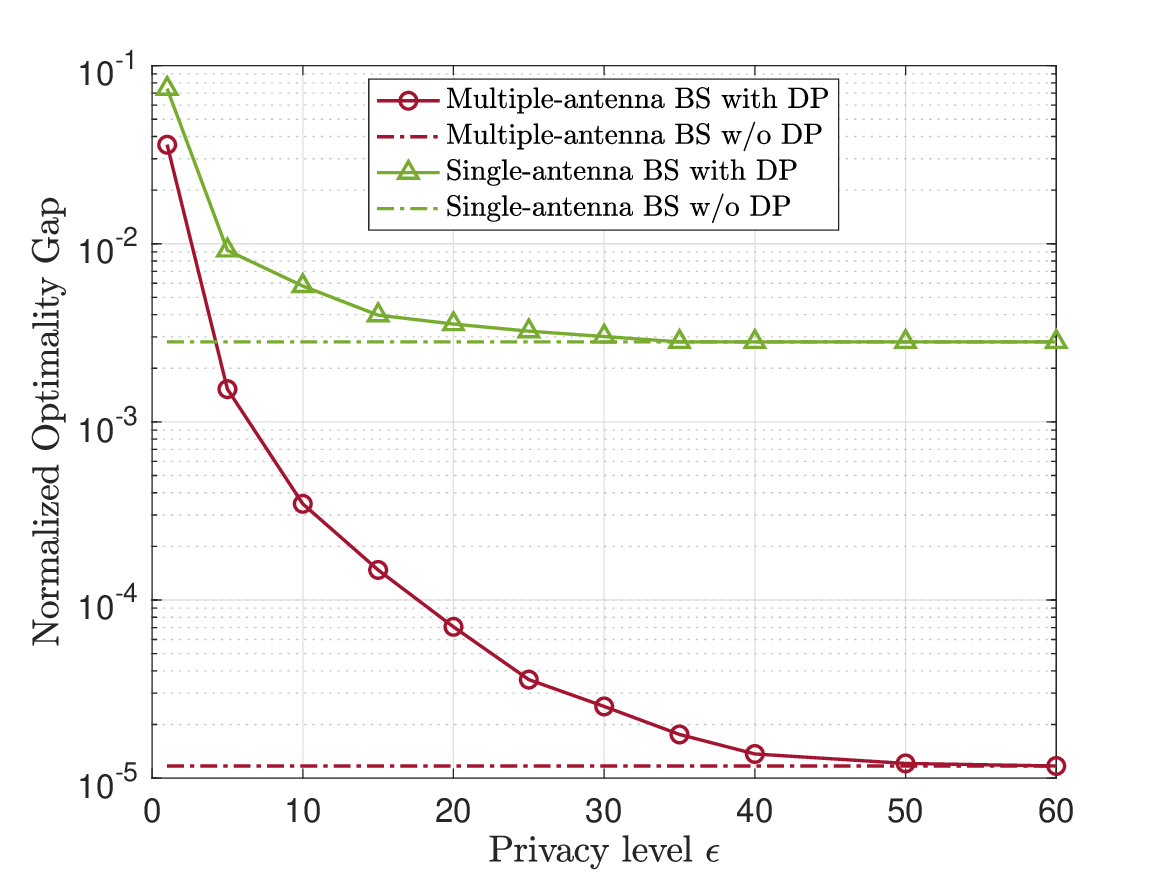}
	\caption{Normalized optimality gap of ridge regression versus the value of $\epsilon$. A small $\epsilon$ corresponds to a more stringent privacy requirement.}
	\label{fig3}
	%\end{minipage}
\end{figure}
\begin{figure}[!t]
	%\begin{minipage}[t]{0.49\linewidth}
	\centering
	\includegraphics[width=2.8 in]{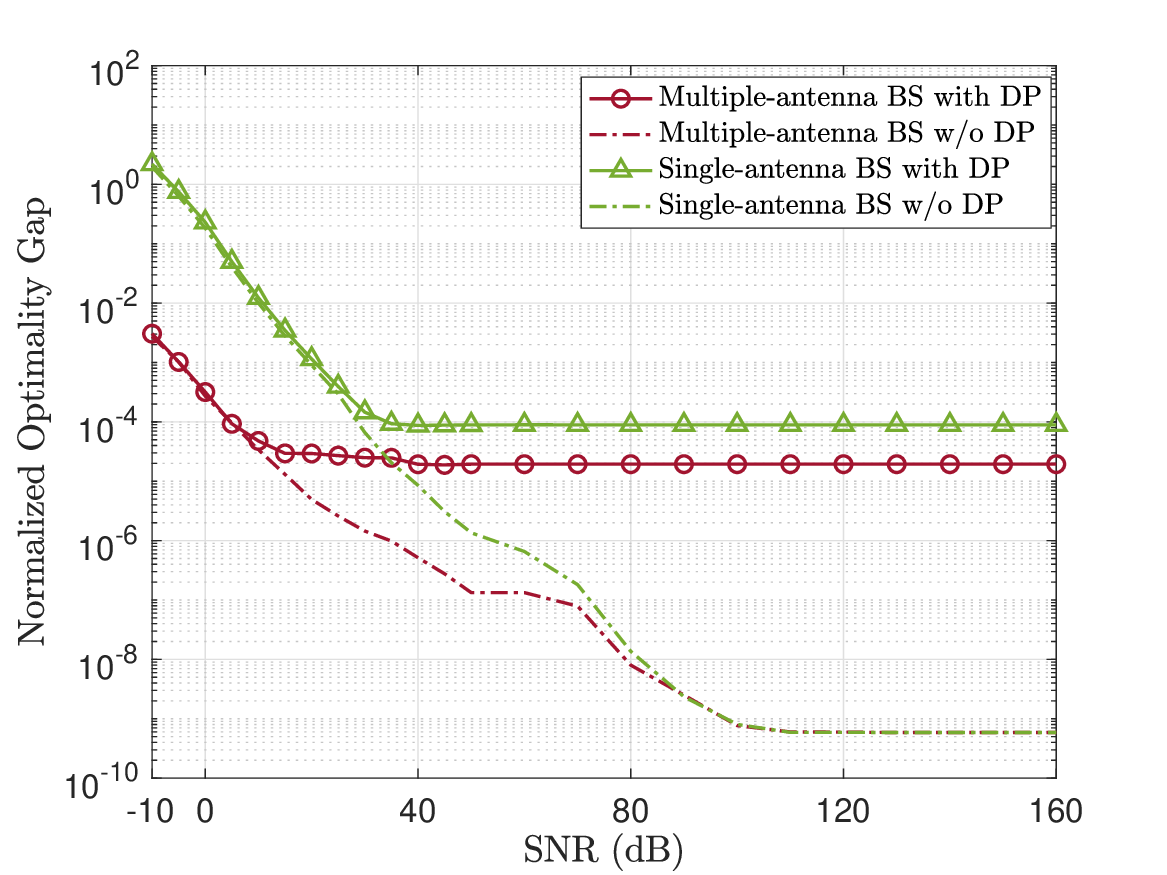}
	\caption{Normalized optimality gap versus SNR with $\epsilon=30$.}
	\label{fig4}
	%\end{minipage}%
\end{figure}
Fig. \ref{fig3} plots the normalized optimality gap versus different values of $\e$. Unless otherwise specified, the number of BS antennas is set to $N=20$ for the schemes with a multiple-antenna receiver, and the SNR is $15$ dB.
Here, a smaller $\e$ represents a higher privacy requirement for the WDs and results in more DP-preserving artificial noise. Fig. \ref{fig3} shows a learning-privacy trade-off: A better learning performance is obtained with a larger $\e$. Moreover, the training loss of the proposed DP-enhanced FL system converges to that without DP as $\e$ increases.  Additionally, the scheme with a multiple-antenna receiver outperforms its single-antenna counterpart, as the proposed receive combining technique achieves a more accurate model aggregation result.

Next, we study the impact of transmit SNR on the training performance in Fig.  \ref{fig4} with a fixed privacy level $\e=30$. On one hand, in the low-SNR regime, the learning accuracy of the DP-enhanced schemes approaches that of the DP-free baselines, as the inherent communication noise is substantial enough to preserve privacy, and therefore, artificial noise is not required. On the other hand, as the communication noise diminishes in the high-SNR regime, additional artificial noise is necessary to maintain privacy. As a result, the training loss 
of the DP-enhanced schemes does not decrease in the high-SNR regime.

\begin{figure}[!t]
	%\begin{minipage}[t]{0.49\linewidth}
	\centering
	\includegraphics[width=2.8 in]{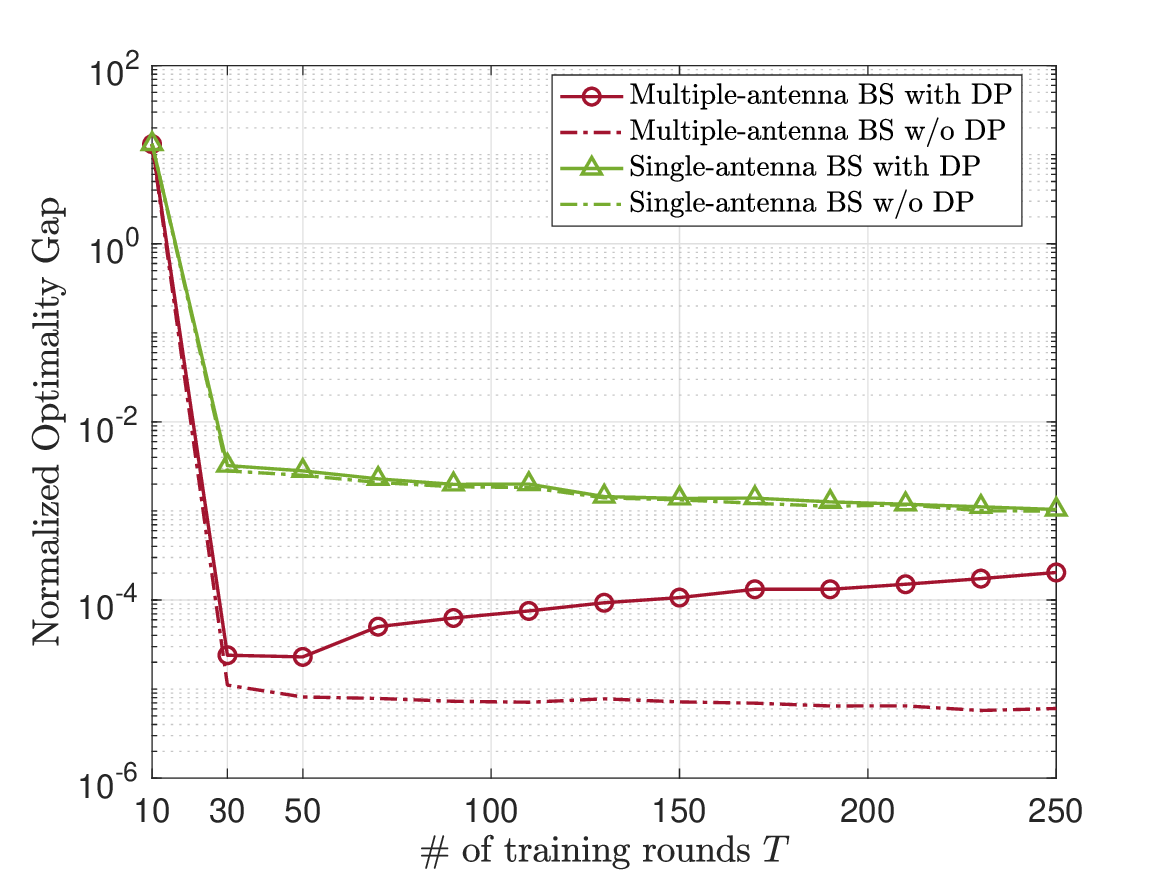}
	\caption{Normalized optimality gap versus the number of training rounds $T$ with $\epsilon$ $=30$ and SNR$=15$ dB.}
	\label{fig5}
\end{figure}
%\end{minipage}
%\begin{minipage}[t]{0.49\linewidth}
\begin{figure}[!t]
	\centering
	\includegraphics[width=2.8 in]{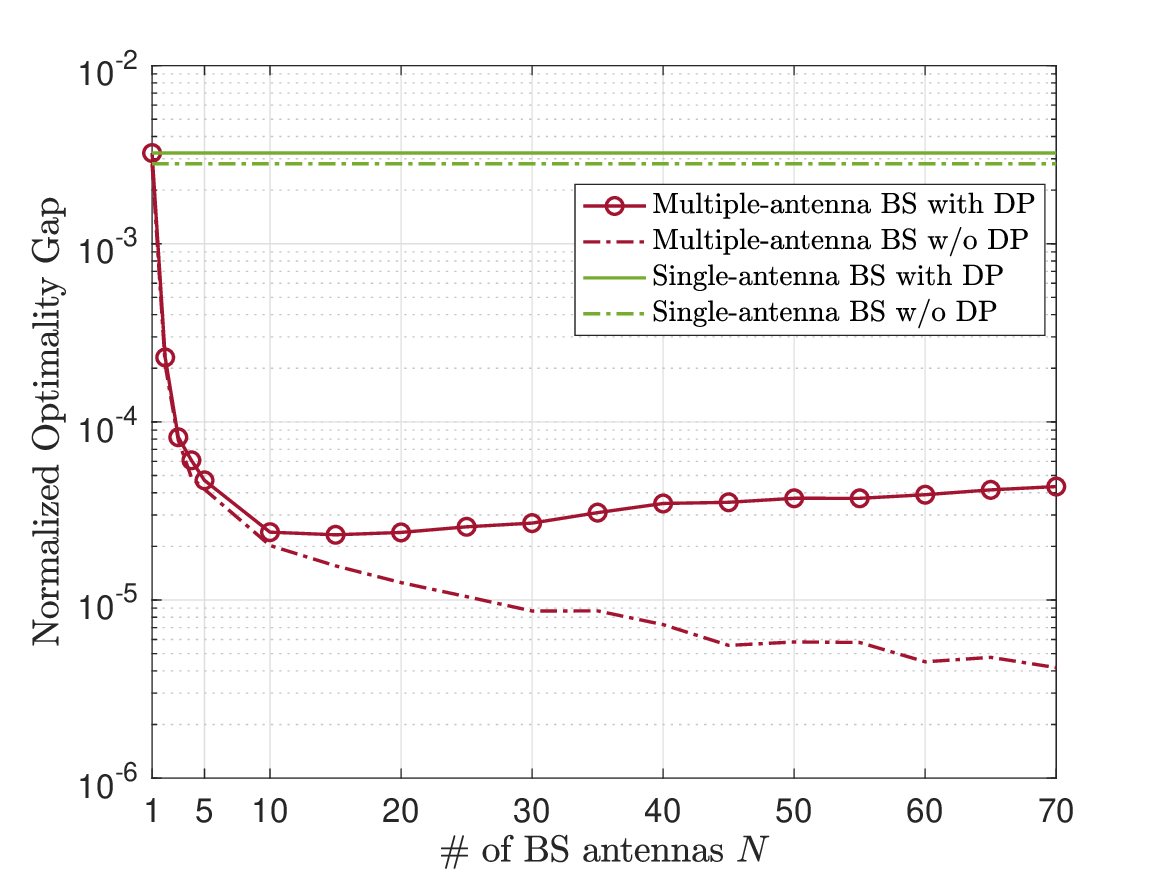}
	\caption{Normalized optimality gap versus the number of BS antennas $N$ with $\e=30$ and $T=30$.}
	\label{fig6}
	%\end{minipage}%
\end{figure}
%\begin{figure}[!t]
%	\begin{minipage}[t]{0.49\linewidth}
%		\centering
%		\includegraphics[width=3.4 in]{figure/heter.eps}
%		\caption{Normalized optimality gap under different data heterogeneity level ${K_1}/{K}$. A larger $K_1/K$ means a more heterogeneous data distribution.}
%		%		\label{figa}
%	\end{minipage}
%	\begin{minipage}[t]{0.49\linewidth}
%		\centering
%		\includegraphics[width=3.4 in]{figure/M.eps}
%		\caption{Normalized optimality gap versus the number of WDs $M$ with uniform data distribution.}
%		%		\label{figb}
%	\end{minipage}%
%\end{figure}
We investigate the impact of the number of training rounds $T$  in Fig. \ref{fig5}. For the baseline without DP, a larger $T$ results in a smaller training loss due to more model aggregation steps. In contrast, the effect on the DP-aware schemes is not always positive; see the red solid curve in  Fig. \ref{fig5}.  When $T$ is small, increasing $T$ significantly improves the training performance, as more gradient descent steps are taken to update the model. However, with artificial noise added in every training round for privacy preservation, the gradient errors are accumulated over iterations. Therefore, further increasing $T$ leads to a larger optimality gap when $T>50$.
This observation aligns with the analytical result in Theorem \ref{theorem1}.
As a conclusion,  the choice of $T$ in FL systems with DP is crucial, especially with multiple-antenna receivers, as $T$ balances the trade-off between learning performance and privacy.

Fig. \ref{fig6} plots the optimality gap versus the number of BS antennas $N$. The baseline with a single-antenna BS, corresponding to $N=1$, is also included for comparison. We see a non-monotonic impact of $N$ on the training performance. Increasing the number of receive antennas from $N=1$  improves the model aggregation accuracy and thus reduces the training loss. However, as $N$ increases, the information extractors at the BS become more powerful, requiring the WDs to add more artificial noise and resulting in a larger optimality gap. 
Moreover, the multiple-antenna scheme generally achieves a better learning-privacy trade-off than its single-antenna counterpart, as the former achieves a smaller training loss under the same privacy requirement.  

\begin{figure}[!t]
	\centering
	\includegraphics[width=2.8 in]{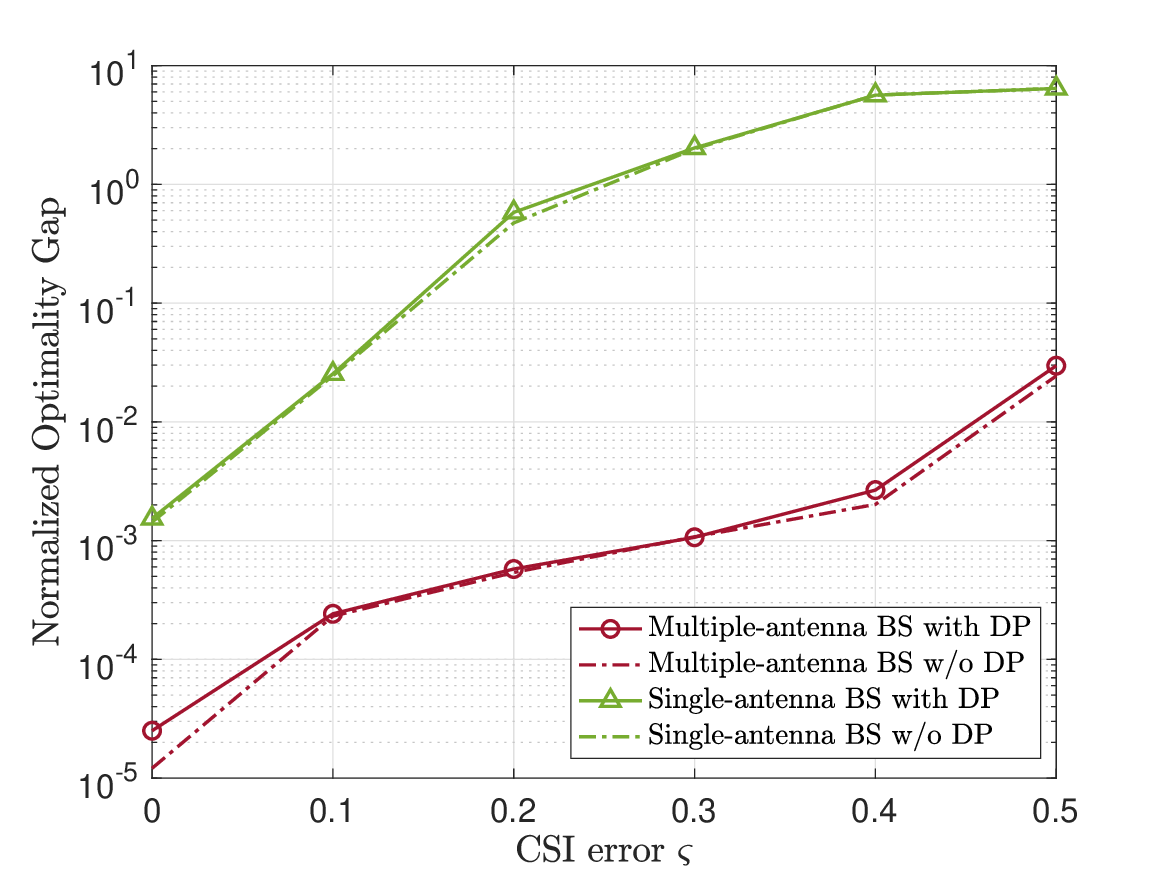}
	\caption{FL performance of the ridge regression task versus the CSI error level $\varsigma$.}
	\label{figc}
\end{figure}

{
	Finally, we study the impact of CSI error on the system optimization. Here, we assume that the transceiver variables are optimized using channel estimates, denoted by $\{\widehat \hv_m\}$, instead of the actual channels $\{\hv_m\}$. The CSI error is characterized by the additive error model as
	\begin{align}
		\widehat \hv_m=\sqrt{\varsigma}\hv_m+\sqrt{1-\varsigma} \ev_m,\forall m,
	\end{align}
	where $\ev_m\sim\CN({\bf 0},\Iv_N)$ denotes the channel estimation error independent of the true coefficients, and $\varsigma\in [0,1]$ characterizes the degree of the CSI error. A larger $\varsigma$ means greater errors in channel estimation. Fig. \ref{figc} plots the learning performance versus the value of $\varsigma$ with $T=30$ and $\epsilon=30$. We see that great CSI errors introduce significant mismatches in over-the-air model aggregation, consequently widening the optimality gap. Moreover, in the presence of CSI errors, the perturbation resulting from channel errors becomes the dominant factor, overshadowing the artificial noise introduced at the transmitters. As a result, the algorithms both with and without DP exhibit comparable accuracies.
}
\subsection{Experiments on Image Classification}\label{sec_6d}
\begin{figure}[!t]
	%\begin{minipage}[t]{0.49\linewidth}
	\centering
	\includegraphics[width=2.8 in]{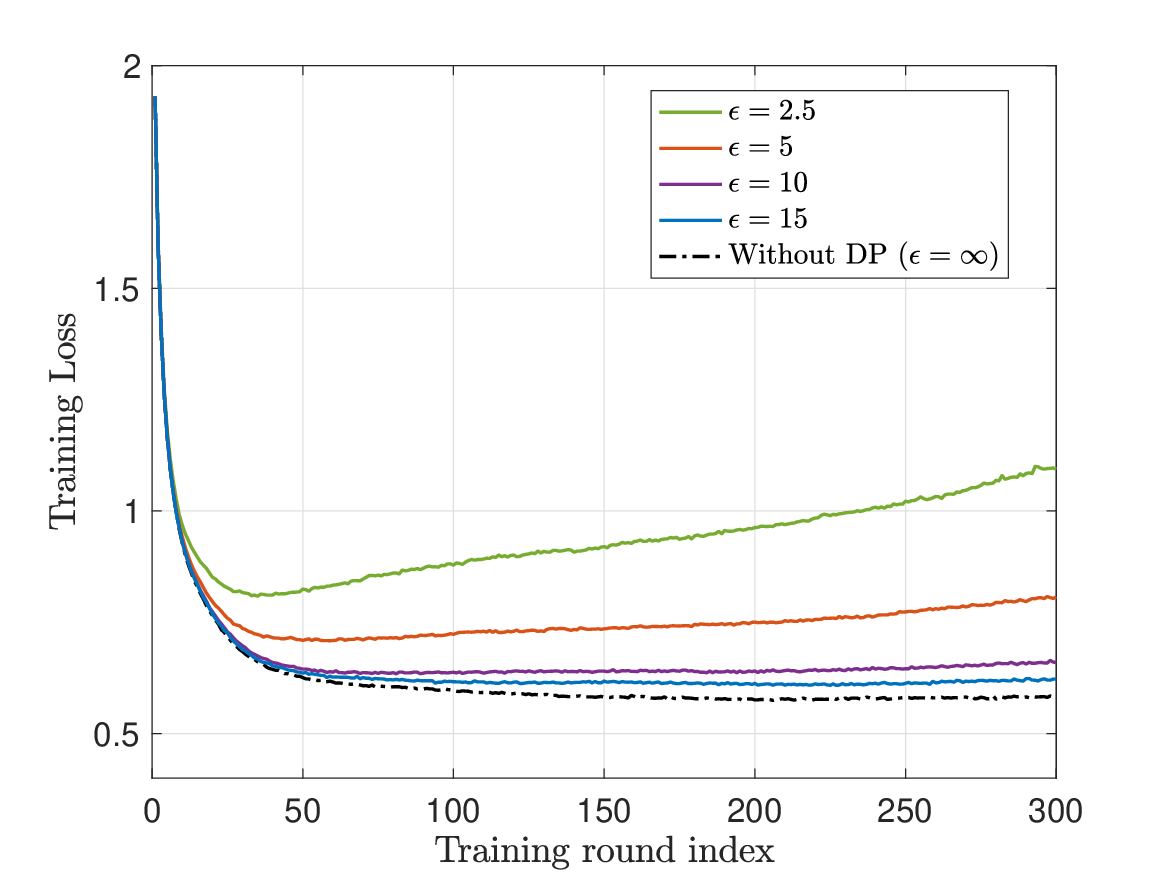}
	\caption{Training loss of image classification over the Fashion-MNIST dataset under different DP levels $\epsilon$ with $T=300$.}
	\label{fig7}
\end{figure}
\begin{figure}[!t]
	%\end{minipage}
	%\begin{minipage}[t]{0.49\linewidth}
	\centering
	\includegraphics[width=2.8 in]{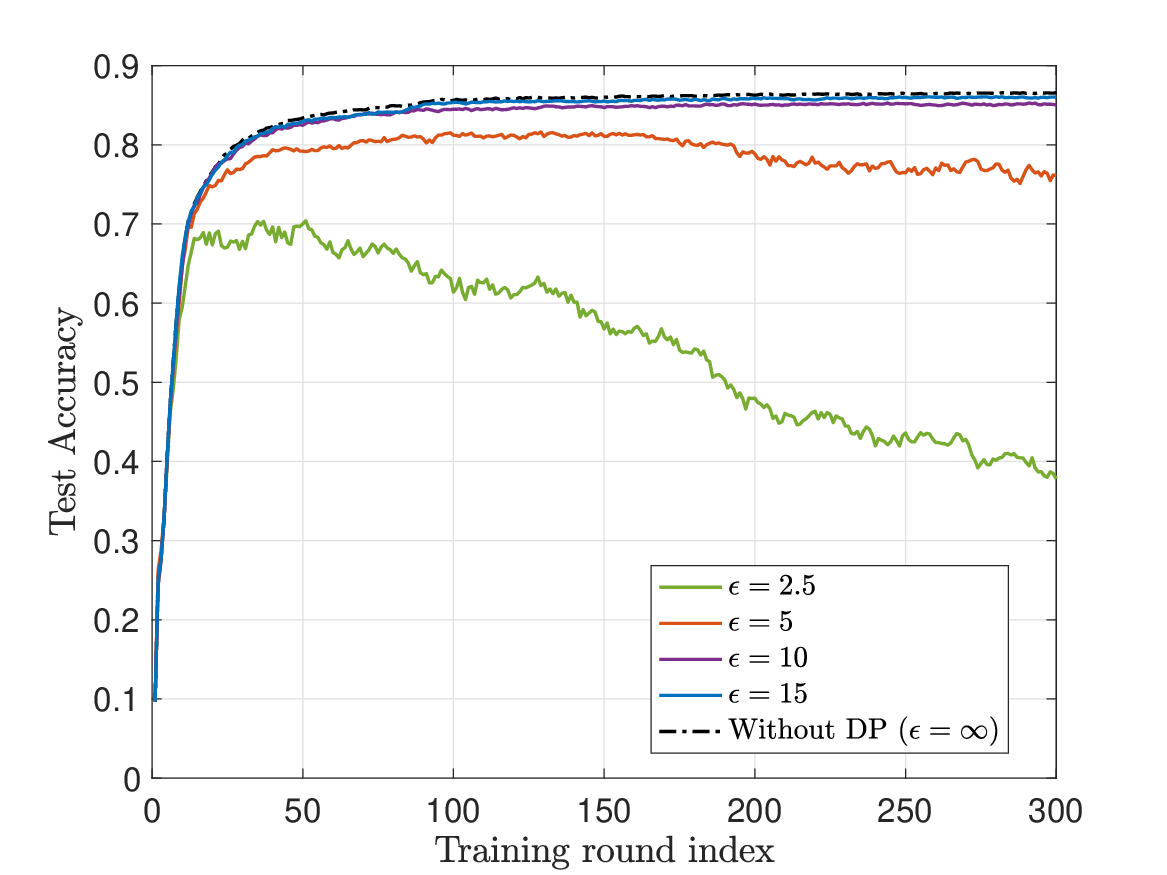}
	\caption{Test accuracy on the Fashion-MNIST dataset under different DP levels with $10,000$ test samples.}
	\label{fig8}
	%\end{minipage}%
\end{figure}
\begin{figure}[!t]
	%\begin{minipage}[t]{0.49\linewidth}
	\centering
	\includegraphics[width=2.8 in]{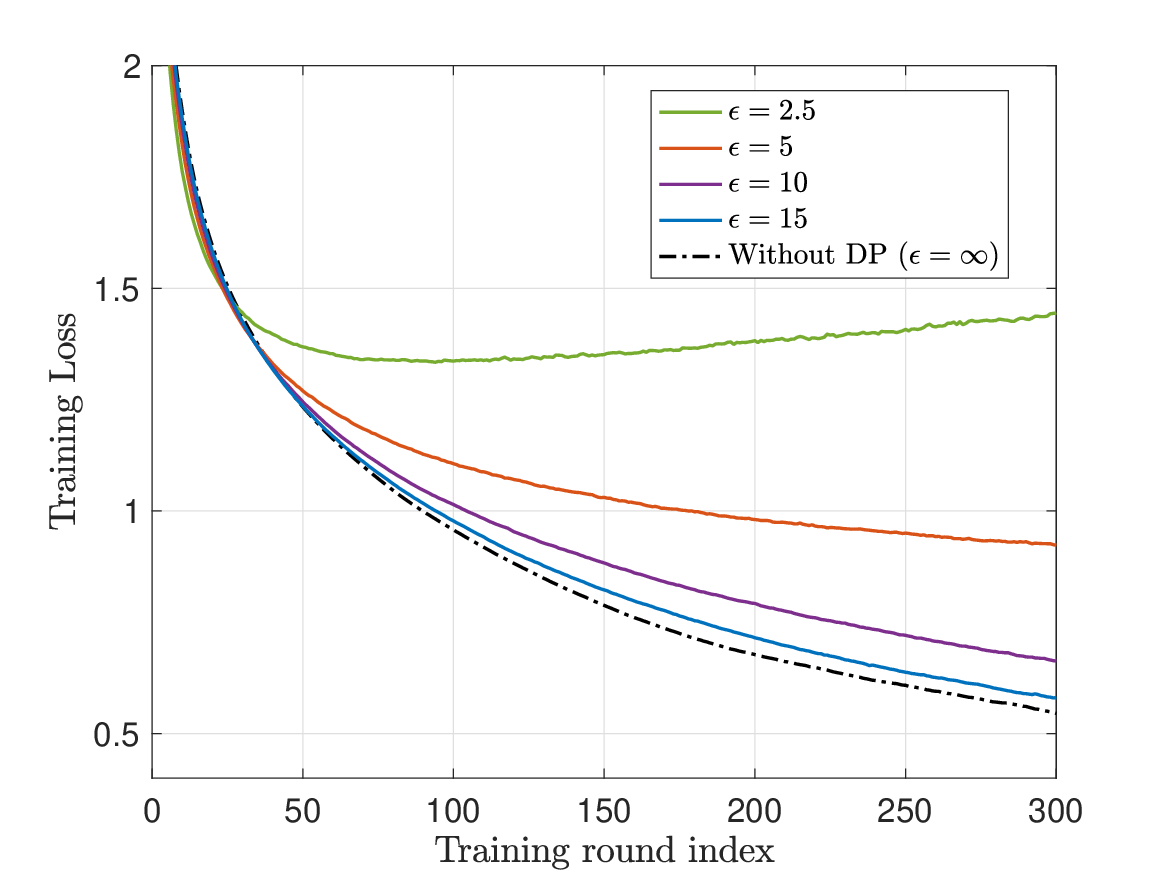}
	\caption{Training loss on the CIFAR-10 dataset under different DP levels $\epsilon$ with $T=300$.}
	\label{fig9}
\end{figure}
\begin{figure}[!t]
	%\end{minipage}
	%\begin{minipage}[t]{0.49\linewidth}
	\centering
	\includegraphics[width=2.8 in]{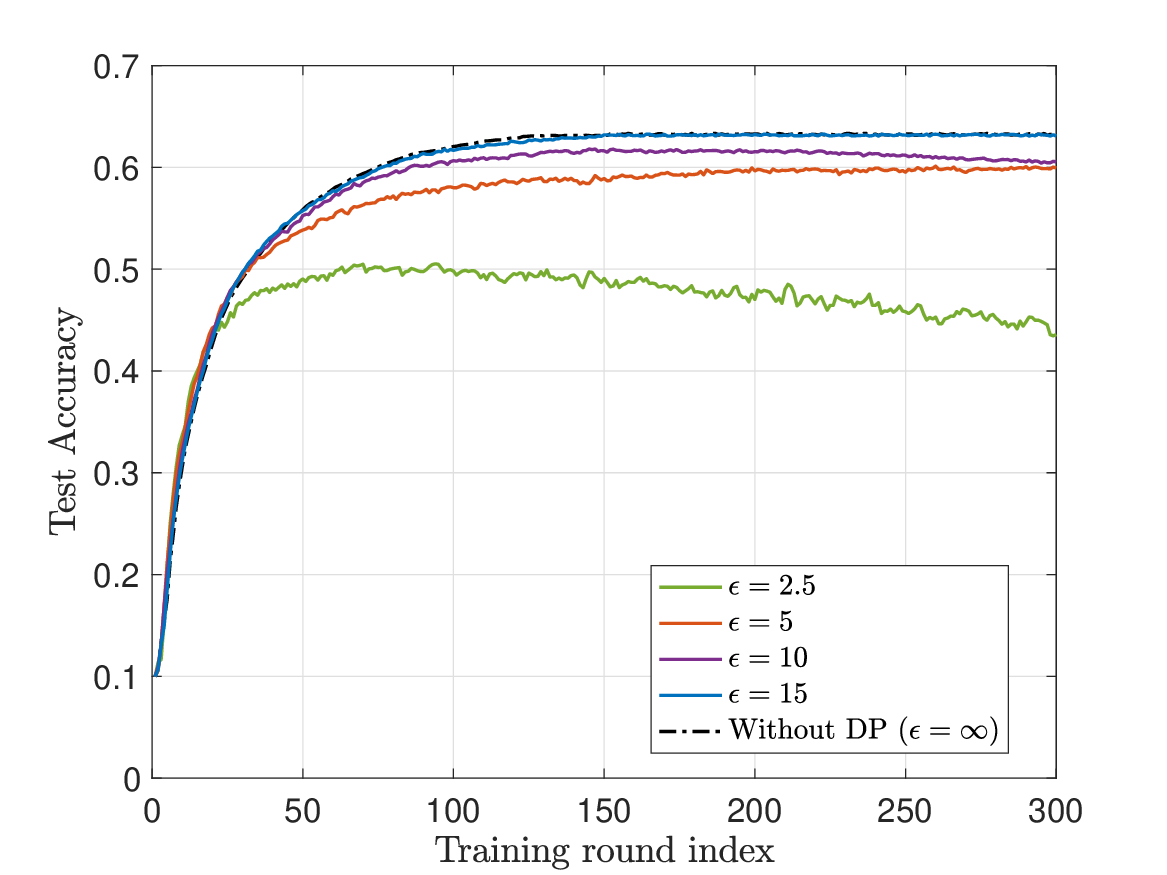}
	\caption{Test accuracy on the CIFAR-10 dataset under different DP levels.}
	\label{fig10}
	%\end{minipage}%
\end{figure}
In this section, we study the FL learning accuracy of the proposed algorithm, \ie Algorithm \ref{alg1}, on image classification with CNNs trained on the Fashion-MNIST \cite{Xiao2017} and CIFAR-10 datasets \cite{Cifar}.  

In the first experiment, $30,000$ training samples from the Fashion-MNIST dataset are uniformly distributed over $M=10$ WDs. The studied CNN architecture is given in \cite[Section V-A]{liu2020reconfigurable} with $d=21,921$ parameters. The loss function is the cross-entropy loss between the model outputs and the true labels. Local training is conducted with mini-batch SGD with a learning rate of $\lambda=0.01$ and a batch size of $128$. We evaluate the training performance by the training loss as well as the test accuracy, which is the ratio of correctly classified test images to the total number of $10^4$ test images (in the range of $[0,1]$). 
We set $T=300,N=20,\text{SNR}=15$ dB, and $\epsilon_m^{\text{(WD)}}=\e,\forall m$. Figs. \ref{fig7} and \ref{fig8} plot the training loss and test accuracy, respectively, under different privacy levels $\e$. Similar to Fig. \ref{fig3}, for an undemanding privacy requirement, \eg $\e\geq 15$, the training accuracy approaches that of the baseline without DP.  Decreasing $\e$ leads to a more stringent privacy requirement and a lower test accuracy.  For a very restricted DP demand, i.e., $\e\leq 2.5$, the artificial noise overwhelms the local gradients, making the training iteration diverge.   

{
	Moreover, we simulate the training loss and the test accuracy of training a CNN on the CIFAR-10 dataset in Figs. \ref{fig9} and \ref{fig10}. Here, we uniformly partition $50,000$ training samples over $M=10$ WDs. We adopt the CNN model described in \cite{FEDSGD} with $d=62,071$ model parameters. For local training, we use mini-batch SGD to minimize the cross-entropy loss with a learning rate of $\lambda=0.005$ and a batch size of $64$. The other system parameters are the same as those in Fig. \ref{fig7}. The results show a consistent privacy-learning trade-off with the experiments on the Fashion-MNIST dataset. Decreasing $\epsilon$ indicates a tighter privacy constraint, resulting in reduced test accuracy. Specifically, with loose privacy constraints (i.e., $\epsilon\geq 15$), the training accuracy aligns closely with the baseline without considering DP. In contrast, when privacy demands become stricter (for instance, at $\epsilon=2.5$), the training process tends to diverge.
}
\section{Conclusions}\label{sec7}
In this work, we addressed the transceiver design problem for differentially private over-the-air FL over a MIMO fading channel. We analyzed the convergence rate and the piracy loss, based on which the optimization problem was formulated as a DP-constrained convergence rate maximization problem. We demonstrated that an edge server with multiple receive antennas can take advantage of receive combining as separate private information extractors to amplify the privacy loss of individual gradient vectors. Furthermore, we proved that the optimal linear information extractor can be calculated by the MMSE estimator of an uplink multi-user system with Gaussian inputs. Consequently, model aggregation over a MIMO channel is more susceptible to privacy leakage, and the zero-artificial-noise mechanism is generally sub-optimal in this setup. To tackle this challenge, we proposed an iterative algorithm based on alternating optimization for transmit scaling and receive beamforming. Numerical results demonstrate a better learning-privacy trade-off of the proposed approach compared with the existing baselines. 
\appendices
\section{Proof of Theorem \ref{theorem1}}\label{appa}
Recall that we update the global model in the $t$-th iteration by
\begin{align}
	\wv_{t+1}=\wv_{t}-\frac{\lambda}{K}\hat{\mathbf{g}}_t=\wv_{t}-\lambda(\nabla F(\wv_{t})-\mathbf{e}_t),
\end{align}
where $\nabla F(\wv_{t})$ is the noiseless gradient of $F(\wv)$ at $\wv=\wv_{t}$, and $\mathbf{e}_t$ is the gradient error vector defined as
%\begin{align}
$	\mathbf{e}_{t}\triangleq({\mathbf{g}_{t}-\hat \gv_t})/{K}$.
%\end{align}

With Assumptions 1-4 and $\lambda=\frac{1}{\omega}$,  we have the following inequality by \cite[Lemma 2.1]{Friedlander2012}
\begin{align}\label{main_ineq}
	\mathbb{E}[F(\wv_{t+1})]
	%	&\leq \mathbb{E}[F(\wv_t)]-\frac{1}{2\omega}\|\nabla F(\wv_{t})\|_2^2+\frac{1}{2\omega}\E[\|\mathbf{e}_{t}\|_2^2]\nonumber \\
	&	\leq \left( 1-\frac{\mu}{\omega}\right) \mathbb{E}[F(\wv_t)-F(\wv^\star)]+\frac{1}{2\omega}\E[\|\mathbf{e}_{t}\|_2^2],
\end{align}
where the expectations are taken w.r.t. the communication noise.

Then, we expand $\frac{1}{2\omega}\mathbb{E}[\|\mathbf{e}_{t}\|_2^2]$ as \eqref{e2}, shown on top of the next page.
\begin{figure*}
	\begin{align}\label{e2}
		\frac{1}{2\omega}\mathbb{E}[\|\mathbf{e}_{t}\|_2^2]&=\underbrace{\frac{1}{2\omega K^2}\sum_{i=1}^d\mathbb{E}\left[\left|\sum_{m=1}^M\left(K_m-\frac{\fv_0^H\hv_m{s_{m,1}}}{\sqrt{\eta}L}\right)g_{m,t}[i]\right|^2\right]}_{C_{t}(\{s_{m,1}\},\eta,\fv_0)}
		+\underbrace{\frac{1}{2\omega K^2}\sum_{i=1}^d\mathbb{E}\left[\frac{\sum_{m=1}^M\left|\fv_0^H\hv_m{s_{m,2}}n_{m,t}[i]\right|^2+\left|\fv_0^H\zv_t[i]\right|^2}{\eta}\right]}_{	A(\{s_{m,2}\},\eta,\fv_0)}.
	\end{align}
	\hrulefill
\end{figure*}
%
%where $A$ and $C_t$ are defined in \eqref{A} and \eqref{C}, respectively.
Combining \eqref{main_ineq} and \eqref{e2}, we have
\begin{align}\label{main_ineq3}
	\mathbb{E}[F(\wv_{t+1})-F(\wv^\star)]\leq B\mathbb{E}[F(\wv_t)-F(\wv^\star)]+{A+C_t}.
\end{align}
By applying \eqref{main_ineq3} recursively with $T$ iterations, we have \eqref{upperbound}.
\section{Proof of Proposition \ref{pro1}}\label{appb}

Since $	C_{t}(\{s_{m,1}\},\eta,1)\geq 0$, for any $\eta$ and $\{s_{m,2}\}$ the  objective in \eqref{eq61a} is lower bounded by
\begin{align}\label{eqaapb1}
	\eqref{eq61a}\geq A(\{s_{m,2}\},\eta,1)\frac{1-B^{T}}{1-B},
\end{align}
where the equality holds if and only if
%
%\begin{align}
$K_m-\frac{h_ms_{m,1}}{\sqrt{\eta}L}=0, \forall m$.
%\end{align}
%
As a result, the optimal solution to $\{s_{m,1}\}$ must satisfy 
\begin{align}\label{eq66}
	s_{m,1}=\frac{\sqrt\eta LK_m\bar h_m}{|h_m|^2}, \forall m.
\end{align}

With $\{s_{m,1}\}$ in \eqref{eq66}, the problem in \eqref{eq61} can be recast as
\begin{subequations}\label{eq68}
	\begin{align}
		\min_{\eta,\{s_{m,2}\}}&~~\frac{\sum_{m=1}^M|h_m|^2|s_{m,2}|^2+\sigma_z^2}{\eta},\label{eq68a}\\
		\text{s.t. }& \eta\leq \frac{1}{  TL^2}\min_m\frac{\sum_{m=1}^M |h_m|^2|s_{m,2}|^2+\sigma_z^2}{\varphi_m} ,\label{eq68b}\\
		& \eta\leq \frac{1}{L^2}\min_m\frac{|h_m|^2}{K_m^2}\left(P_{\text{max}}-|s_{m,2}|^2 \right).\label{eq68c}
	\end{align}
\end{subequations}
%%
%
%Since $|s_{m,2}|^2\geq 0$, the conditions in \eqref{eq68b}--\eqref{eq68c} also imply
%\begin{subequations}\label{eq69}
%	\begin{align}
%	&	\eta\leq \frac{\sigma_z^2}{T\max_m\varphi_m}  , \\
%	&	
%	\end{align}
%\end{subequations}. 
We discuss the optimal solution to \eqref{eq68} in two situations.
\begin{enumerate}
	\item  When $T< T_0$, using the definition of $T_0$, we have
	\begin{align}
		&\min_m	\frac{|h_m|^2}{K_m^2}\left(P_{\text{max}}-|s_{m,2}|^2 \right)\leq P_{\text{max}}\min_m\frac{|h_m|^2}{K_m^2}\nonumber\\&\leq \min_m\frac{\sigma_z^2}{T\varphi_m} 
		\leq\min_m\frac{\sum_{m=1}^M |h_m|^2|s_{m,2}|^2+\sigma_z^2}{T\varphi_m}.
	\end{align}
	In other words, the constraint in \eqref{eq68c} always implies \eqref{eq68b}. Accordingly, \eqref{eq68} is equivalent to minimizing \eqref{eq68a} under \eqref{eq68c}. Note that for any $\{s_{m,2}\}$, the objective in \eqref{eq68a} is inversely proportional to $\eta$, and thus the optimal $\eta$ is given by the maximum feasible $\eta$ under \eqref{eq68c}, \ie
	\begin{align}\label{temp1}
		\eta=\frac{1}{L^2}\min_m\frac{|h_m|^2(P_{\text{max}}-|s_{m,2}|^2)}{K_m^2}.
	\end{align}
	Plugging \eqref{temp1} into \eqref{eq68}, the remaining optimization problem is 
	%
	%	\begin{align}\label{temp2}
	$\min_{\{s_{m,2}\}}\frac{\sum_{m=1}^Mh_m^2|s_{m,2}|^2+\sigma_z^2}{\min_m\frac{|h_m|^2(P_{\text{max}}-|s_{m,2}|^2)}{K_m^2}}$.
	%	\end{align}
	%
	It can be verified that the optimal solution is given by $s_{m,2}=0, \forall m$. Substituting  $s_{m,2}=0$ into \eqref{temp1}, we have \eqref{case2a}.
	
	\item When $T\geq T_0$, we have the following two sub-cases:
	\begin{enumerate}
		\item If $\{s_{m,2}\}$ satisfies $\frac{\sum_{m=1}^Mh_m^2|s_{m,2}|^2+\sigma_z^2}{T\max_{m}\varphi_m}\leq \min_m\frac{|h_m|^2(P_{\text{max}}-|s_{m,2}|^2)}{K_m^2}$,  the constraint \eqref{eq68b} always implies \eqref{eq68c}. On one hand, optimizing \eqref{eq68} is equivalent to minimizing   \eqref{eq68a} under \eqref{eq68b}. Similar to \eqref{temp1}, the optimal $\eta$ is given by
		\begin{align}\label{temp3}
			\eta=\frac{\sum_{m=1}^M|h_m|^2|s_{m,2}|^2+\sigma_z^2}{L^2T\max_{m}\varphi_m}.
		\end{align}
		On the other hand, we see by plugging \eqref{temp3} into \eqref{eq68} that any feasible $\{s_{m,2}\}$
		%		 satisfying $\frac{\sum_{m=1}^Mh_m^2|s_{m,2}|^2+\sigma_z^2}{T\max_{m}\varphi_m}\leq \min_m\frac{|h_m|^2(P_{\text{max}}-|s_{m,2}|^2)}{K_m^2}$ 
		is optimal to \eqref{eq68a} with the objective value given by $L^2T\max_{m}\varphi_m$.
		\item If $\frac{\sum_{m=1}^Mh_m^2|s_{m,2}|^2+\sigma_z^2}{T\max_{m}\varphi_m}> \min_m\frac{|h_m|^2(P_{\text{max}}-|s_{m,2}|^2)}{K_m^2}$,  the constraint \eqref{eq68c} always implies \eqref{eq68b}. Therefore, the  optimal $\eta$ is given by \eqref{temp1}.  Accordingly, the objective value is lower bounded by  $\text{\eqref{eq68a}}> L^2T\max_{m}\varphi_m$, showing that $\{s_{m,2}\}$ in this case is not globally optimal.
	\end{enumerate}  
	By combining the above two scenarios 1) and 2), we complete the proof of Proposition \ref{pro1}. 
\end{enumerate}
{\section{Proof of the Convergence of Algorithm \ref{alg1}}\label{appc}
	Recall that Algorithm \ref{alg1} solves \eqref{eq44} iteratively to output a sequence of solutions $\{\{s_{m,2}(i)\},\eta(i),\fv_0(i)\}$, where $i=1,2,\cdots,$ denotes the iteration index in the outer loop of Algorithm \ref{alg1}, \ie the value of $\mbox{iter}$ in Step 3. Our goal is to prove that the sequence $\{A(\sv(i),\eta(i),\fv_0(i))\}_{i=1}^\infty$ converges to a stationary value of $A(\cdot)$ as $i\to\infty$.

	For ease of notation, we denote $\sv(i)=[s_{1,2}(i),\cdots,s_{M,2}(i)]$ and the output of the inner loop in Step 6 as $\Fv^{(j)}(i)$ and $\tau^{(j)}(i)$, where $j$ is the inner-loop iteration index. The objectives in \eqref{eqsub2} and \eqref{eqsub3} are represented as $h_i(\Fv,\tau)$ and $g_i(\Fv,\tau|\Fv^{(j)}(i))$, respectively, and the feasible set in \eqref{eq44} is denoted by $\mathcal{F}$. Finally, we note that the variable transformation from $\fv_0$ and $\eta$ to $\Fv$ and $\tau$ in \eqref{eqsub1} is bijective and continuous. 
	
	The first step is to show $\{A(\sv(i),\eta(i),\fv_0(i))\}_{i=1}^\infty$ converges, \ie the limit $\lim_{i\to \infty}A(\sv(i),\eta(i),\fv_0(i))$ exists.
	To begin, note that the objective $A(\cdot)$ is uniformly bounded below given that $A(\cdot)\geq 0$. 
	Further, using \eqref{eqsub3} and following the argument in \cite[Eq. (4)]{7547360}, we have
	\begin{align}
		A&(\sv(i-1),\eta(i-1),\fv_0(i-1))=A(\sv(i-1),\tau^{(0)}(i),\Fv^{(0)}(i))\nonumber\\
		&\geq A(\sv(i-1),\tau^{(1)}(i),\Fv^{(1)}(i))
		\geq \cdots \nonumber\\
		&\geq A(\sv(i-1),\tau^{(\infty)}(i),\Fv^{(\infty)}(i))
		=	A(\sv(i-1),\eta(i),\fv_0(i)).
	\end{align}
	Given that $\sv(i)$ is the minimizer of \eqref{eq46}, we have
	\begin{align}
		A(\sv(i),\eta(i),\fv_0(i))\leq A(\sv(i-1),\eta(i-1),\fv_0(i-1)).
	\end{align}
	%This confirms that the sequence $\{A(\sv(i),\eta(i),\fv_0(i))\}_{i=1}^\infty$ is non-increasing. 
	As a result, the non-increasing and bounded nature of this sequence implies the convergence of the objective.
	
	Next, we prove that $\lim_{i\to \infty}A(\sv(i),\eta(i),\fv_0(i))$ corresponds to a stationary value  for some $\rho\geq 0$ in \eqref{eqsub2}. To proceed, we introduce the following assumption:
	%  \assumption{\label{ass5}The regularization hyper-parameter $\rho$ in the DC step \eqref{eqsub2} is properly selected such that \eqref{eqsub1} and \eqref{eqsub2} are equivalently framed: In essence, a stationary point of one problem is also a stationary point to the other.}
	\assumption{\label{ass6}In each iteration, the principal eigenvector of $\Fv^{(j)}(i)$ can be uniquely calculated by the eigenvalue decomposition. This translates to the condition that the principal eigenvalue of $\Fv^{(j)}(i)$ is singly multiplicative. }
	\lemma{
		Under Assumption \ref{ass6}, the following conclusions hold:
		\begin{enumerate}
			\item[R1] $g_i(\Fv,\tau|\Fv^{(j)}(i))$, given by the objective in \eqref{eqsub3}, is continuous with both $(\Fv,\tau)$ and $\Fv^{(j)}(i)$.
			\item[R2] $g_i(\Fv,\tau|\Fv^{(j)}(i))$ is continuously differentiable with
			respect to $(\Fv,\tau)$.
			\item[R3] $h_i(\Fv,\tau)$, given by the objective in \eqref{eqsub2}, is continuously differentiable with
			respect to $(\Fv,\tau)$.
			
			\item[R4] For any $i$, given an the initial value $h_i(\Fv^{(0)}(i),\tau^{(0)}(i))<\infty$, the sublevel set, denoted by $\text{lev~} h_i\triangleq \{\tau,\Fv\in \mathcal{F}|h_i(\Fv,\tau)\leq h_i(\Fv^{(0)}(i),\tau^{(0)}(i))\}$, is compact.
		\end{enumerate}
	}
	\begin{proof}
		From \eqref{eqsub3}, $g_i(\Fv,\tau|\Fv^{(j)}(i))$ is a linear function with respect to $(\Fv,\tau)$, ensuring its continuity and differentiability with $(\Fv,\tau)$. Moreover, $g_i(\Fv,\tau|\Fv^{(j)}(i))$ is related to the principal eigenvector of $\Fv^{(j)}(i)$. Under Assumption 2, this relation is continuous with $\Fv^{(j)}(i)$. Moreover, R3 follows from that $h_i(\cdot)$ is a linear function.
		
		Regarding R4, by the Heine–Borel theorem \cite{MathAna}, the compactness of the sublevel set can be obtained once it is proven to be both closed and bounded.  We first note that the continuity of $h_i(\cdot)$ implies the closeness of $\text{lev~}h_i$. To see this, for any convergent sequence $\{\Fv_t,\tau_t\}\in \text{lev~}h_i$, its limit also belongs to $\text{lev~}h_i$ due to the continuous nature of $h_i(\cdot)$.
		Meanwhile, the coerciveness\footnote{A function $h(\xv)$ is coercive if $\|\xv\|\to\infty$ implies $h(\xv)\to\infty$.} of $h_i(\cdot)$ over its feasible set, combined with the finiteness of the initial objective value, confirms the boundedness of $\text{lev~}h_i$. This completes the proof.
	\end{proof}
	Applying Lemma 1 to \cite[Section II-C]{7547360}, we conclude that the outer loop of Algorithm \ref{alg1} after the MM steps converges to a stationary value. Specifically, the objective value $h_i(\Fv^{(\infty)}(i),\tau^{(\infty)}(i))$ in \eqref{eqsub2} converges to its stationary value. Since the objective in \eqref{eqsub2} can be taken as the Lagrangian function of that in \eqref{eqsub1}, $(\Fv^{(\infty)}(i),\tau^{(\infty)}(i))$ is also a stationary point to the objective in \eqref{eqsub1} for some optimal choice of $\rho$.
	Given the continuous transformation from $(\Fv,\tau)$ to $(\fv_0,\eta)$, we confirm that for any given $\sv(i-1)$,
	\begin{align}\label{eq63}
		\nabla_{\fv_0,\eta}A(\sv(i-1),\eta(i),\fv_0(i))={\bf 0}.
	\end{align}
	
	On the other hand, as $\sv(i)$ solves the linear programming problem in \eqref{eq46}, we have
	\begin{align}\label{eq64}
		\nabla_{\sv}A(\sv(i),\eta(i),\fv_0(i))={\bf 0}.
	\end{align}
	
	Using \eqref{eq63} and \eqref{eq64} and the convergence of the objective value, $\{A(\sv(i),\eta(i),\fv_0(i))\}_{i=1}^\infty$ converges to a stationary value as $i\to\infty$.
}

\begin{footnotesize}
	\bibliographystyle{IEEEtran}
	\bibliography{IEEEabrv,ref}
\end{footnotesize}

\end{document}